\documentclass[prb,showpacs,twocolumn,floatfix]{revtex4}
\usepackage{graphicx}
\begin{document}
\def\rhov{{\mbox{\boldmath{$\rho$}}}}
\def\tauv{{\mbox{\boldmath{$\tau$}}}}
\def\Lambdav{{\mbox{\boldmath{$\Lambda$}}}}
\def\sigmav{{\mbox{\boldmath{$\sigma$}}}}
\def\xiv{{\mbox{\boldmath{$\xi$}}}}
\def\oh{{\scriptsize 1 \over \scriptsize 2}}
\def\of{{\scriptsize 1 \over \scriptsize 4}}
\def\tf{{\scriptsize 3 \over \scriptsize 4}}
\title{Towards a Microscopic Model of Magnetoelectric Interactions in
${\bf Ni_3V_2O_8}$}

\author{A. B. Harris,$^1$ T. Yildirim,$^2$ A. Aharony,$^{3,4}$ and
O. Entin-Wohlman,$^{3,4}$}

\affiliation{
(1) Department of Physics and Astronomy, University of Pennsylvania,
Philadelphia, PA 19104\\(2) NIST Center for Neutron
Research, National Institute of Standards and Technology,
Gaithersburg, Maryland 20899\\ (3) School of Physics and
Astronomy, Raymond and Beverly Sackler Faculty of Exact Sciences,
Tel Aviv University, Tel Aviv 69978, Israel\\(4) Department of
Physics, Ben Gurion University of the Negev, Beer Sheeva 84105, Israel}
\date{\today}

\begin{abstract}
We develop a microscopic magnetoelectric coupling in Ni$_3$V$_2$O$_8$
(NVO) which gives rise to
the trilinear phenomenological coupling used previously to explain
the phase transition in which magnetic and ferroelectric order
parameters appear simultaneously.  Using combined neutron scattering
measurements and first-principles calculations of the phonons in NVO,
we determine eleven phonons which can induce the observed spontaneous
polarization. Among these eleven phonons, we find that a few of them
can actually induce a significant dipole moment. Using the calculated atomic
charges, we find that the required distortion to induce the observed
dipole moment is very small (~0.001 \AA) and therefore it would be
very difficult to observe the distortion  by neutron-powder diffraction.
Finally, we identify the derivatives of the
exchange tensor with respect to atomic displacements which are
needed for a microscopic model of a spin-phonon coupling in NVO
and which we hope will be obtained
from a fundamental quantum calculation such as LDA+U.  
We also analyze two toy
models to illustrate that the Dzyaloskinskii-Moriya interaction
is very important for coexisting of magnetic and ferroelectric
order but it is not the only mechanism when the local site
symmetry of the system is low enough.
\end{abstract}
\maketitle

\section{Introduction}

Recently studies have identified
a family of multiferroics which display a phase transition
in which there simultaneously develops long-range incommensurate
magnetic and uniform ferroelectric order. Perhaps the most detailed
studies have been carried out on the systems Ni$_3$V$_2$O$_8$
(NVO)\cite{Rogado,LawesKenzelmann,FERRO,EXPT} and
TbMnO$_3$ (TMO).\cite{TMO1,TMO2} (For a review, see Ref. \onlinecite{REV}.)
This phenomenon has been explained\cite{FERRO} on
the basis of a phenomenological model which invokes a Landau
expansion in terms of the order parameters describing the
incommensurate magnetic order and the order parameter describing
the uniform spontaneous polarization.  Already from this
treatment it was clear that a microscopic model would have to
involve a trilinear interaction Hamiltonian proportional to the
product of two spin variables and one displacement variable.
Furthermore, the symmetry requirements of the phenomenological model
would naturally be realized by a proper microscopic model.
Accordingly, in this paper we present a detailed combined neutron
scattering study and first-principles calculations of the 
optical phonons of NVO and thereby identify those having the right
symmetry to induce a dipole moment.  We also consider the expansion of the
exchange tensor to first order in the generalized displacement coordinates.
The aim of this programme is to determine which of the generalized
displacements are relevant and which corresponding elements of the
exchange tensor are needed for this microscopic calculation.
This paper will therefore set the stage for a separate quantum
calculation of the exchange tensor and its derivatives with respect
to atomic displacements.  On general grounds, one might expect the
Dzyaloshinskii-Moriya\cite{DM1,DM2} (DM) interaction to play an
important role in this calculation,\cite{DAGOTTO} 
and indeed, we will find this to be the case.
However, the conclusion is a bit more subtle, in that
displacement derivatives of many elements of the exchange
tensor (and not just the antisymmetric one of the DM
interaction) are needed for the microscopic calculation.
We will show that simplistic models are somewhat misleading
in that they may lead one to believe that the only exchange tensor
elements whose derivatives are important are the DM ones.
Although our calculations may seem complicated, they
are essential if one wishes to actually relate the
magnetoelectric coupling to detailed quantum calculations
(via LDA\cite{LDA} or LDA$+U$\cite{LDAU} schemes) of the strain-dependence
of the exchange tensors.  The methodology of the present paper can
be extended in a straightforward way to TMO, for instance.

Briefly this paper is organized as follows.  In Sec. II we 
give an overview of the calculation. 
In Sec. III we discuss the first-principles calculations of
the zone-center phonons and identify those phonons  which
transform like a vector and which are thus candidates to
produce a spontaneous polarization. In this section we
also present the neutron scattering measurements of the phonon
density of states (DOS), which is found to be in good agreement
with the calculated spectrum.  Next in Sec. IV we
summarize the results of the determination of the spin
structure.\cite{LawesKenzelmann,EXPT} In Sec. V we then use the
symmetry operations of the crystal to show how the phonon
derivatives of the various exchange tensors in the unit cell
are related to one another.  Then in Sec. VI we show that a mean-field
treatment of this spin-phonon coupling leads to the results
obtained previously\cite{LawesKenzelmann,EXPT} in a phenomenological
model.  Here we give expressions for the spontaneous polarization
in terms of graidents of the excahnge tensor whose evaluation
remains to be done.  It would be nice to have a simple model to illustrate these
results.  However, our studies of two ``toy models'' in Sec. VII
indicate that they do not reproduce some essential features of our
complete calculation.  Finally, our conclusions are summarized
in Sec. VIII.

\section{Overview of Calculation}

Here we give a brief qualitative overview of the calculation.
First we review the symmetry
of the space group of NVO, Cmca (No. 64 in Ref. \onlinecite{HAHN}).
The space group operations (apart from primitive translations) are
specified in Table \ref{PUC}.  (Here and below, sites within the unit cell
are given as fractions of the sides of the {\it conventional} unit cell,
so that $(x,y,z)$ denotes $(xa,yb,zc)$.)  We now describe the sets of
crystallographically equivalent sites which the various atoms occupy.
(Such a set of crystallographically equivalent
sites is called a {\it Wyckoff orbit}.)  There are six such orbits as
shown in Table~\ref{orbit}.  The first two are those of the Ni atoms,
the first consisting of the two Ni(1) (a) sites (which we call
``cross-tie'' sites) and the second consisting of the four
Ni(2) (e) sites (which we call ``spine'' sites).  The four V (f) sites
comprise the third orbit and the oxygen sites are distributed into
two (f) orbits, one containing four O(1) atoms, the other containing four
O(2) atoms, and a (g) orbit containing eight O(3) atoms. (The letters
a, e, f, g classify the site symmetry according to the convention of
Ref. \onlinecite{HAHN}. The number of sites in the orbit as
listed in Ref. \onlinecite{HAHN} is twice what we give here because here
we consider the primitive unit cell rather than the conventional
unit cell.)  The locations of these sites are specified in the second
column of Table \ref{orbit}.
Note that there are two formula units of NVO per unit cell.
The Ni sites form buckled planes which resemble a kagom\'{e}
lattice and two such adjacent planes are shown in Fig. 1 where
the buckling is omitted for simplicity. There one sees that the
Ni spine sites form chains along the $x$-direction between which
the Ni cross-tie sites are situated, with bonds to nearest-neighboring
spine sites which form a cross tie.
To illustrate the use of Table \ref{orbit} we find that the  eight
operations of Table \ref{PUC} acting on $(0,0,0)$
generate four copies of each of the 2 a sites
which are at $(0,0,0)$ and $(1/2,0,1/2)$.  Similarly one can
generate the eight g sites by applying in turn the eight operations
of Table \ref{PUC} to the site at $(x,y,z)$. (In each case it may be
necessary to bring the site back into the original unit cell via
a primitive translation vector.)

\begin{table}
\begin{tabular} {||c | c ||}
\hline
${\rm E}{\bf r} =(x,y,z)$ 
& $2_z{\bf r} =(\overline x + 1/2, \overline y , z + 1/2)$ \\
$2_y {\bf r} = (\overline x + 1/2, y, \overline z +1/2)$
& $2_x {\bf r} = (x, \overline y , \overline z)$ \\
${\cal I} {\bf r} =(\overline x,\overline y,\overline z)$ 
&$m_z {\bf r} =(x+\oh , y, \overline z + \oh )$ \\
$m_y {\bf r} = (x+\oh , \overline y, z +1/2)$
&$m_x{\bf r} = (\overline x, y , z)$ \\
\hline \end{tabular}
\caption{\label{PUC}General positions within the primitive unit cell for
Cmca which describe the symmetry operations of this space group.
$2_\alpha$ is a two-fold rotation (or screw) axis and
$m_\alpha $ is a mirror (or glide) plane. 
The primitive translation vectors are ${\bf a}_1=(a/2)\hat i + (b/2)\hat j$,
${\bf a}_2=(a/2) \hat i - (b/2)\hat j$, and ${\bf a}_3=c\hat k$,
where $a=5.92170\AA$, $b=11.37105\AA$, and
$c=8.22638\AA$.\protect{\cite{EXPT,SAUER}}}
\end{table}

\begin{table}
\begin{tabular}{|c|c|c|c|}\hline\hline
Atoms   &  (x/a,y/b,z/c) & Wyckoff  &  Decomposition \\ \hline
Ni(1)   & (0,0,0)      &   2a    & A$_u$+2B$_{1u}$+2B$_{2u}$+B$_{3u}$ \\ \hline
Ni(2)   & (1/4,y/b,1/4)      &   4e   & A$_u$+2B$_{1u}$+B$_{2u}$+2B$_{3u}$ \\ 
        & y=0.1298 (0.1304)     &     & A$_{g}$+2B$_{1g}$+B$_{2g}$+2B$_{3g}$\\ \hline
V(1)    & (0,y,z)      &    4f & A$_u$+2B$_{1u}$+2B$_{2u}$+B$_{3u}$ \\ 
        & y=0.3762 (0.3762)     &     & 2A$_{g}$+2B$_{1g}$+B$_{2g}$+2B$_{3g}$\\ 
	& z=0.1197 (0.1196)     &     &                                        \\ \hline
O(1)    & (0,y/b,z/c)      &    4f & A$_u$+2B$_{1u}$+2B$_{2u}$+B$_{3u}$ \\ 
        & y=0.2481 (0.2490)     &     & 2A$_{g}$+2B$_{1g}$+B$_{2g}$+2B$_{3g}$\\ 
	& z=0.2308 (0.2301)     &     &                                        \\ \hline	
O(2)    & (0,y/b,z/c)      &    4f & A$_u$+2B$_{1u}$+2B$_{2u}$+B$_{3u}$ \\ 
        & y=0.0011(0.0008)     &     & 2A$_{g}$+2B$_{1g}$+B$_{2g}$+2B$_{3g}$\\ 
	& z=0.2444 (0.2441)     &     &                                        \\ \hline
O(3)    & (x/a,y/b,z/c)      &    8g & 3A$_u$+3B$_{1u}$+3B$_{2u}$+3B$_{3u}$ \\ 
        & x=0.2656(0.2703)     &     & 3A$_{g}$+3B$_{1g}$+3B$_{2g}$+3B$_{3g}$\\ 
	& y=0.1192 (0.1184)     &     &                                        \\ 
	& z=0.0002 (0.0012)     &     &                                        \\ \hline \hline
\end{tabular}
\caption{\label{orbit}
In column 2 we give the Wyckoff positions, the fractional
coordinates (in coumn 2) and their multiplicity  (in column 3)
of atoms in the Wyckoff orbits of NVO listed in column 1.
In column 2 we give the values of the structural parameters
(e.g. $x$, $y$, and $z$) as deduced from
diffraction data\protect{\cite{EXPT,SAUER}}
and the corresponding values we find
from the structural minimization are given in parentheses.  In
column 4 we give the symmetry decomposition of the displacements
of atoms in each of the Wyckoff orbits.}
\end{table}

We now review briefly the nature of the ordered phases which occur
as the temperature $T$ is lowered at zero external magnetic
field.\cite{Rogado,LawesKenzelmann,EXPT}
At high temperatures the system is paramagnetic.
When $T$ is lowered through the value $T_{PH} \approx 9.1$K,
an incommensurate phase appears (which we call the high-temperature
incommensurate or HTI phase) in which the Ni spins on
the spine chains are oriented very nearly along the $x$-axis  with
a modulation vector also along $\hat i$.  (The axes are denoted
either ${\bf a}$, ${\bf b}$, and ${\bf c}$, or $x$, $y$, and $z$ and
corresponding unit vectors are denoted $\hat i$, $\hat j$, and $\hat k$.)
As the temperature is further
lowered through the value $T_{HL}\approx 6.3$K transverse order
appears at the same incommensurate wavevector and also order
appears on the cross-tie sites, as shown in Fig. 1. We call
this phase the low-temperature incommensurate or LTI phase.  Within
experimental uncertainty, these two ordering transitions are
continuous.  As the temperature is lowered through the value
$T_{LC} \approx 4$K, a discontinuous transition occurs in which
a commensurate antiferromagnetic phase appears.  In this
phase antiferromagnetism results from the arrangement of
spins within the unit cell in such a way that the magnetic
unit cell remains identical to the paramagnetic unit cell.

\begin{figure}
\includegraphics[height=7.0cm]{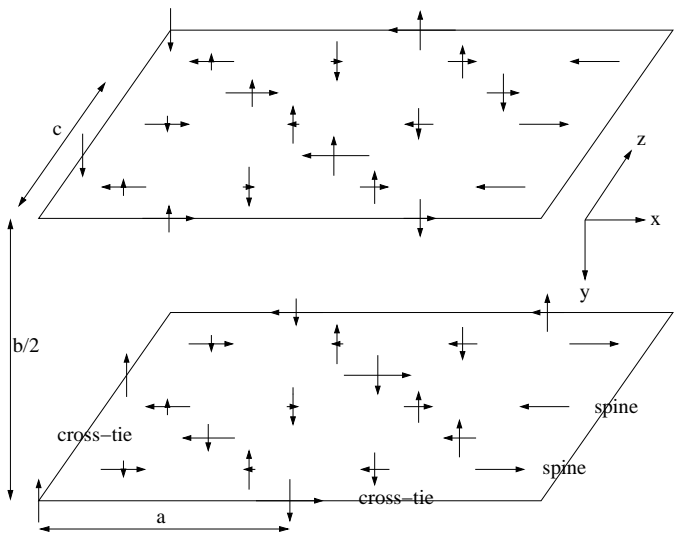}
\caption{\label{SSTR} Schematic diagram showing
the $x$- and $y$-components of the spins in the LTI
phase. We used the parameters: $q=0.28(2 \pi /a)$,
$a_{s,x}=1.6$, $a_{c,y}=1.4$, $b_{s,y}=1.3$, $b_{c,x}=-2.2$, and
$\phi_L=\phi_H + \pi/2$.  The small $z$-components of spin
are not represented.
The planes are buckled, so that alternately spine chains are displaced
above and below the planes shown (but this buckling is not shown).
In the HTI phase the cross-ties have negligible moments and the spine
chains have the incommensurately modulated longitudinal moments similar
to those shown. The a spin components are odd under $2_x$ and
the b spin components are even under $2_x$, where $2_x$ is a
two-fold rotation about the $x$ axis.}
\end{figure}

Perhaps surprisingly, it was found that ferroelectricity coincides with
the existence of LTI order, and this behavior was explained by
a phenomenological model based on a Landau expansion
in powers of order parameters describing ferroelectricity
and  those needed to describe the magnetic ordering of the
HTI and LTI phases.\cite{FERRO}
The phenomenological interaction $V$ between the magnetic and
ferroelectric order parameters was given as
\begin{eqnarray}
V&=& \sum_{\alpha} \sum_{X,Y=L,H} a_{\alpha, X,Y} \sigma_X({\bf q})
\sigma_Y (-{\bf q}) P_\alpha \ ,
\label{VINT} \end{eqnarray}
where $\sigma_{X}({\bf q})$ is the  order parameter describing
incommensurate order at wavevector ${\bf q}$ characteristic of
the HTI phase (for X=H) or for the additional ordering
appearing in the LTI phase (for X=L) and
$\alpha$ labels the Cartesian component of the uniform spontaneous
polarization vector ${\bf P}$. Using the symmetry properties of
the order parameters $\sigma_H({\bf q})$ and $\sigma_L({\bf q})$ it was
shown that only $a_{b,L,H}$ and $a_{b,H,L}$ are nonzero. This result
provides a phenomenological explanation for the experimental finding
that a nonzero polarization is induced by incommensurate magnetism
only in the LTI phase and then only with {\bf P} along the {\bf b} axis.

This phenomenological theory elucidates the symmetry of the
magnetically induced ferroelectric state.  To develop an
analogous microscopic theory, one could treat a Hubbard-like Hamiltonian
involving the 3d electrons of the Ni ions and also the 2p electrons
of the O ions.  From such a treatment one can obtain the
{\it spin Hamiltonian} which describes the ``low energy''
sector of this Hubbard-like Hamiltonian.  This low-energy
sector is obtained by removing states whose energy relative
to the low energy sector involves Coulomb integrals.  In this
approach one develops a canonical transformation to eliminate
hopping matrix elements.  This approach is obviously most
appropriate for insulators.  This type of calculation was
formalized by Anderson\cite{PWA} who thereby obtained the 
antiferromagnetic Heisenberg model from a superexchange mechanism.
More recently this approach has been applied
to LaCu$_2$O$_4$\cite{TY,OE} and LaTiO$_3.$\cite{RS1,RS2} 
Here we will use the phenomenological model to deduce the form
of the low energy spin Hamiltonian which describes
magnetically induced ferroelectricity in NVO.  We are
currently analyzing the more basic Hubbard-like Hamiltonian
to show how it gives rise to the spin Hamiltonian.

From the form of Eq. (\ref{VINT}), it is clear that the spin-phonon
Hamiltonian we seek must be of the form
\begin{eqnarray}
V_{\rm sp-ph} &=& \sum_{i,j,k} \sum_{\alpha \beta \gamma}
b_{\alpha \beta \gamma_k} S_\alpha (i) S_\beta (j) Q_{\gamma_k} \ ,
\label{MICRO}
\end{eqnarray}
where ${\bf S}(i)$ is the vector spin operator for site $i$ and
$Q_{\gamma_k}$ is the $k$th normal mode amplitude at zero
wavevector which transforms like the $\gamma$ component of a
first rank tensor (vector).  We will discuss the normal modes in some
detail in the next section.  To implement the interaction of
Eq. (\ref{MICRO}) it is convenient to classify both the normal modes
and the spin components according to their transformation properties.
This interaction represents a linear potential, {\it i. e.}
a force on the phonon coordinate $Q_{\gamma_k}$.  Up to
quadratic order in the displacements the terms in the elastic
potential energy $V_{\rm el}$ which depend on the $Q_{\gamma_k}$ are
\begin{eqnarray}
V_{\rm el} &=& {1 \over 2} \sum_{\gamma, k}
\omega_{\gamma_k}^2 Q_{\gamma_k}^2 + V_{\rm sp-ph} \ .
\label{HPH} \end{eqnarray}
Minimization of this elastic energy with respect to the phonon
coordinates leads to a phonon displacement which is proportional to
the product of two spin functions, whose symmetry we analyze below.
Furthermore, since the displaced ions carry an
electric charge (albeit an effective charge), these displacements
will give rise to a spontaneous polarization providing the
necessary spin components are nonzero.  This calculation will recover
the symmetry properties of the phenomenological treatment.

\section{Zone-center Phonons; Neutron Scattering Measurements and First 
Principles Calculations}

\subsection{Generalized Displacements}

Since the normal modes are complicated linear combination of atomic
displacements we start by giving a qualitative discussion of
generalized displacements (GD's), which are linear combinations of
atomic displacements which transform according to the various irreducible
representations (irreps), whose characters are given in Table \ref{CHAR},
for the paramagnetic space group of NVO.  The actual phonon modes
are linear combinations (which we will give later) of these basis
functions or GD's.

\begin{table}
\begin{tabular}{||cc|cccccccc|c|}\hline\hline
& &$1$ &  $2_y$ & $2_x$ & $2_z$ & $\cal I$ & $m_y$ & 
$m_x$ & $m_z$ &Function\\
\hline
& $A_g$ & 1 &  1 &  1 &  1 &  1 &  1 &  1 &  1 & x$^2$,y$^2$,z$^2$ \\
& $A_u$ & 1 &  1 &  1 &  1 & -1 & -1 & -1 & -1 & xyz \\
& $B_{2g}$ & 1 &  1 & -1 & -1 &  1 &  1 & -1 & -1 & xz \\
& $B_{2u}$ & 1 &  1 & -1 & -1 & -1 & -1 &  1 &  1 & y\\
& $B_{3g}$ & 1 & -1 &  1 & -1 &  1 & -1 &  1 & -1 & yz \\
& $B_{3u}$ & 1 & -1 &  1 & -1 & -1 &  1 & -1 &  1 & x \\
& $B_{1g}$ & 1 & -1 & -1 &  1 &  1 & -1 & -1 &  1 & xy \\
& $B_{1u}$ & 1 & -1 & -1 &  1 & -1 &  1 &  1 & -1 & z \\
\hline\hline
\end{tabular}
\caption{\label{CHAR}Irreducible representation of the
paramagnetic space group of NVO. The vector representations
are $B_{1u}$, $B_{2u}$, and $B_{3u}$ which transform like
$z$, $y$, and $x$, respectively.}
\end{table}

\begin{figure}
\includegraphics[height=6cm]{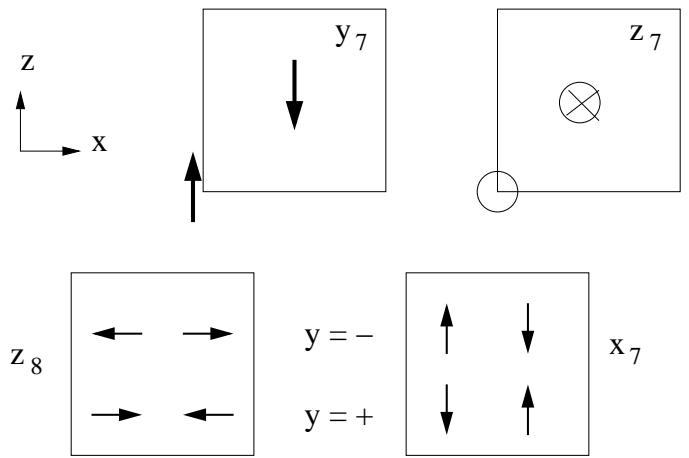}
\caption{\label{GENDIS} Generalized displacements 
$y_7$, $z_7$, $z_8$, and $x_7$ which transform like the
components of a vector, for cross-tie sites (upper panels)
and spine sites (lower panels).  Atomic displacements (for
the GD's indicated by the labels) in the
$x$-$z$ plane are represented by arrows, whereas those in
the $+y$ direction ($-y$ direction) are indicated by
filled (open) circles.
It is easy to see that these modes couple to the uniform 
displacements.  For instance, consider the mode $z_8$ shown
here  and in particular consider how the ionic displacements
of the spine sites which are shown affect
the cross-tie sites (not shown) at
the corners and centers of the square.  Imagine the ion-ion
interactions to be repulsive.  Then the nearest neighbors
at negative $z$ relative to each cross-tie get closer to the
cross tie and the nearest neighbors at positive $z$ relative to
each cross tie get farther away from the cross tie.  Thus, all cross-ties
are squeezed towards positive $z$.  This same reasoning also shows
that even though this motion is confined to the $x$-direction, it will
induce a dipole moment along the $z$-direction.}
\end{figure}

\begin{figure}
\includegraphics[height=3cm]{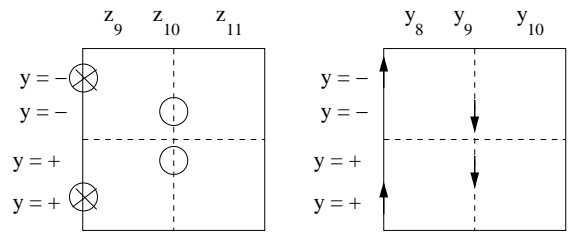}
\caption{\label{GENDIS2} As Fig. \protect{\ref{GENDIS}}.  Here we illustrate
schematically the vector GD's $z_9$, $z_{10}$, $z_{11}$, $y_8$, $y_9$, and
$y_{10}$ for f sites.  The
placement of the sites reproduces the symmetry of an f site and
is not quantitative for either V or O atoms.  Three distinct f sites
are occupied, one by a V atom and the other two by O atoms.} 
\end{figure}

\begin{figure}
\includegraphics[height=5.5cm]{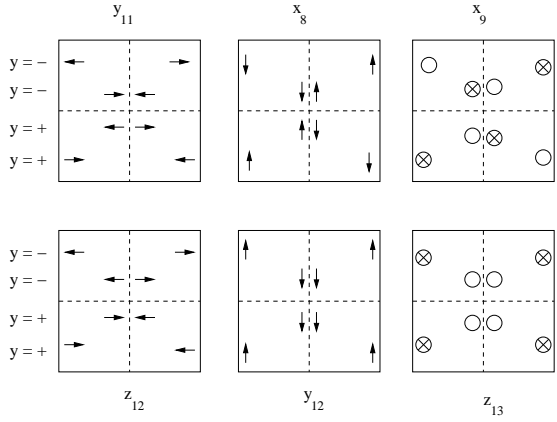}
\caption{\label{GENDIS3} As Fig. \protect{\ref{GENDIS2}}.  Here we illustrate
schematically the vector GD's $y_{11}$, $x_8$, $x_9$, $z_{12}$, $y_{12}$, and
$z_{13}$ for g sites.}
\end{figure}

We are only interested in GD's which transform according to the
vector irreps $B_{nu}$, for $n=1,2,3$.  We now
give a qualitative discussion of these vector GD's.
It is actually quite easy to generate the GD's which
transform like vector components $x$, $y$, and $z$.  First of all,
one sees that assigning all atoms of a given Wykoff orbit the
{\it same} displacement along the $\alpha$-axis will give a GD
which transforms under the symmetry operations of the
space group (given in Table \ref{PUC}) like the $\alpha$ component
of a vector.  Since NVO has 6 crystallographically inequivalent sites
this construction gives six GD's along each of the
three coordinate axes, which we denote $x_n$, $y_n$, and $z_n$, for
$n=1, 2, \dots 6$.  We now discuss the construction of the less trivial
GD's which transform like a vector and which are
shown in Figs. \ref{GENDIS}, \ref{GENDIS2}, and \ref{GENDIS3}.
To generate these results, one can start with
an arbitrarily chosen site to which a vector displacement along one
of the three coordinate axes is specified.  Then one generates the 
displacements of the other sites in the Wykoff orbit 
so as to reproduce the desired transformation properties.
For example, to construct a $z$-like mode on the spines, we could
assign the lower left spine site (in the lower left panel of Fig.
\ref{GENDIS}) a displacement along the $x$-axis.
To be a $z$-like mode the pattern of displacements should be even
under $m_x$, which fixes the displacement of the lower right spine
to be that shown.  Such a $z$-like mode should be odd under a two-fold
rotation about an $x$-axis passing through the center of the cell.
Applying this operation to the two lower spine sites fixes the orientation
of the displacements of the upper spine sites.
The other four symmetry operations give these
same displacements.  Had we started with a spine site with a displacement
along the $y$-axis, we would have gotten a null displacement because
this symmetry with displacements along the $y$-axis is not allowed.  Had
we fixed the first site to have a displacement along the $z$-axis we would
have found the trivial mode in which all sites are displaced in parallel.
The other GD's shown in the figures were generated in
this way.  Of course, the actual normal modes (phonons) consist of linear
combinations of basis functions having the same irrep label and they will
be discussed in the next subsection.
 
\subsection{Normal Modes}

In this section we present inelastic neutron scattering (INS)
measurements of the phonon density of states (DOS) in NVO
along with the first-principles calculations of the zone-center
phonons. We will identify the phonons which have the correct
symmetry to induce a spontaneous polarization. We will also
attempt to estimate the local distortion which gives rise to
the observed dipole moment in NVO.

The INS measurements were performed using the filter analyzer spectrometer (FANS)
located on beamline BT4 at the NIST Center for Neutron Research \cite{fans}.
For energies above 40 meV, a Cu(220) monochromator, surrounded by
$60'-40'$ horizontal collimation and
combined with a cooled polycrystalline beryllium filter analyzer was used.
For the low energy spectrum (i.e. $ E < 40 $ meV), a graphite (PG) monochromator
with $20'-20'$ collimation was used. The relative energy resolution of the FANS
instrument is approximately 5\% in the energy range probed. The powder
Ni$_3$V$_2$O$_8$ sample (about 20 grams) was held at 12 K (paramagnetic phase)
and 8 K (HTI phase) with a helium-filled aluminum can using a closed-cycle
He$^3$ refrigerator. 

The first-principles total-energy and phonon calculations were performed
by the plane-wave implementation of the spin-polarized generalized gradient
approximation (SP-GGA) to density functional theory (DFT).\cite{pwscf}
We used $4\times 4\times 3$ k-points according to Monkhorst-pack scheme
and Vanderbilt ultra-soft pseudopotentials for which a cutoff energy
of 400 eV was found to be enough for total energies to converge within 0.5 eV/atom.
We considered the primitive unit cell of the NVO which contains 26 atoms
as listed in Table~II. Experimental lattice parameters were used in the
calculations but the atomic positions were optimized to eliminate the
forces down to 0.02 eV/\AA. The optimized positions are listed in 
Table.~II, which are in excellent agreement with the experimental positions.
Using the optimized structure, we next calculated the zone-center phonons
and the corresponding INS one-phonon spectrum as described in Ref.
\onlinecite{taner_phonon}.

\begin{figure}
\includegraphics[width=8cm]{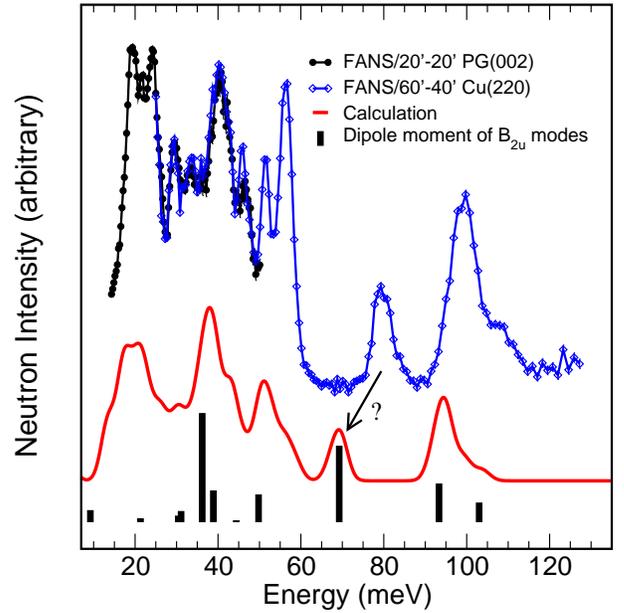}
\caption{\label{ins_dos} The observed and calculated INS spectrum. The black
vertical bars show the B$_{2u}$ phonons whose intensity is proportional
to the induced dipole moment when the system is distorted by the
zero-point rms values of the modes.}
\end{figure}

The measured INS spectrum along with the calculation is shown in Fig.~\ref{ins_dos}.
We observed almost identical spectra in the paramagnetic ($T=12$ K) and
the HTI ($T=8$ K) phases and therefore show only the combined spectrum. 
The agreement of the calculations to the observed spectrum is quite good,
giving further confidence that the first-principle calculations capture
the main physics. It also suggests that the phonon modes in NVO have small
dispersion with wavevector. This is because  the INS spectrum is approximately
averaged over a large range of ${\bf Q}$ and the calculations are only for $Q=0$.
The biggest difference between the INS spectrum and calculation 
is for the observed feature near 80 meV, which is calculated to be around
70 meV. As we discuss  in detail below, interestingly this phonon has
the right symmetry and the atomic displacement vector to induce a large dipole moment.
Hence, maybe the disagreement for the energy of this mode  could be some indication
of strong spin-phonon coupling. 

In order to identify the right phonon modes that can induce the observed
dipole moment along the {\bf b}-axis in NVO, we carried out the symmetry analysis of the
zone center phonons.  Table \ref{CHAR} shows the character table for the
irreducible representations (irreps) of the group $G_{\bf v}$ for optical
phonons at zero wavevector ${\bf v}$. (For a review of group theory see Ref.
\onlinecite{GROUP}.) There are 26 atoms in the primitive unit cell and
the representation $\Gamma_u$ induced by the vector space of these
$26 \times 3 = 78$ atomic displacements has the decomposition
\begin{eqnarray}
\Gamma_u &=& 10 A_g + 8 A_u + 8 B_{1g} + 13 B_{1u} \nonumber \\
&& \ + 7 B_{2g} + 12 B_{2u} + 11 B_{3g} + 9 B_{3u} \ .
\end{eqnarray}
One can check that the vector representations
which transform like $x$, $y$, and $z$ are $B_{3u}$,
$B_{2u}$, and $B_{1u}$, respectively. To discuss the
spontaneous polarization these
are the only irreps we will need to consider.
Among the 78 phonons, twelve have B$_{2u}$ symmetry,
and can therefore produce the observed\cite{FERRO}  spontaneous polarization along
the ${\bf b}$-axis. One of the these twelve modes is acoustic (i.e.
all atoms move uniformly along the {\bf b}-axis) and will not be considered any further.
To calculate the phonon energies and wavefunctions we found the
eigenvalues $\omega^2_n$ of the matrix ${\bf W}(q=0)$, which is related to the
Fourier transform of the potential energy matrix by
$W_{\tau, \alpha ; \tau' , \alpha'}(q=0) = \sqrt{M_\tau} V_{\tau,\alpha;
\tau',\alpha'} (q=0) \sqrt{M_{\tau'}}$, where $\tau$ labels sites within
the primitive unit cell,  $\alpha$ labels Cartesian components, and
$M_\tau$ is the mass of the atom at site $\tau$.  The
corresponding eigenvectors $O_{\tau, \alpha}^{(n)}$ are given in Table \ref{b2u}.
Thus the displacements within the unit cell are given in terms of the
normal modes operators $Q_n$ as
\begin{eqnarray}
u_{\tau, \alpha} &=& \sum_n O_{\tau,\alpha}^{(n)} M_\tau^{-1/2} Q_n \nonumber \\
&=& \sum_n O_{\tau,\alpha}^{(n)} \sqrt{ {\hbar \over 2 M_\tau \omega_n}} 
(a_n^\dagger + a_n ) \ ,
\end{eqnarray}
where $a_n^\dagger$ is a phonon creation operator.  The energies of the
eleven optical modes are shown in Fig.~\ref{ins_dos} by black bars whose height
for mode $n$ is proportional to the average polarization ${\bf P}^{(n)}_y$ of
that mode along the $y$ axis.  The polarization vector of the $n$th mode
is estimated from the following formula:
\begin{equation}
P_{{\rm rms},\alpha}^{(n)} = \frac{1}{\Omega_{\rm uc}} \sum_{\tau} q_\tau 
O_{\tau ,\alpha}^{(n)} Q_{\rm rms} M_\tau^{-1/2} \ ,
\end{equation}
where $\Omega_{\rm uc}$ is the volume of the unit cell and
$Q_{\rm rms}=\sqrt{\hbar/2\omega}$
is the average zero-point fluctuation.  As we can see from
Fig.~\ref{ins_dos}, half of the B$_{2u}$ modes induce relatively small
dipole moments. This is due to the fact that for these
phonons, atoms mainly oscillate along the ${\bf c}$-axis and the
$b$-component of the motion is only a second order effect.
However for the other half, the motion is directly along the
${\bf b}$-axis and therefore the induced dipole moment is significant. 
Animations of these modes and more information can be obtained at
Ref. \onlinecite{taner_web}. We note that two particular phonons, one at 36 meV
and the other around 70 meV, induce a significantly large dipole moment. 
The eigenvectors of these and the other $B_{2u}$ modes along with
the magnitude of the induced dipole moments are given in Table~\ref{b2u}.
\begin{table*}
\begin{tabular}{|c|cc|c|cc|cc|cc|ccc|c|c|c|}\hline\hline
& \multicolumn {2} {|c|} {Ni(1)} & Ni(2) &
\multicolumn{2} {|c|} {V(1)} & \multicolumn {2} {|c|} {O(1)} &
\multicolumn{2} {|c|} {O(2)} & \multicolumn {3} {|c|} {O(3)} & 
$\omega$  & $\frac{Q_{rms}}{\sqrt{m_p}}$& P$_{rms}$  \\ 
 & \multicolumn{2}{|c|} {$q=0.90$e} & $q=0.86$e &
\multicolumn{2} {|c|} {$q=1.18$e} & \multicolumn {2} {|c|} {$q=-0.62$e} &
\multicolumn {2} {|c|} {$q=-0.68$e} & \multicolumn{3} {|c|} {$q=-0.59$e} & 
(meV) & (\AA)& ($10^{-4}$ C/m$^{2}$) \\ \hline\hline
Mode& \multicolumn {2} {|c|} {$(0,y,z)$} & $(0,y,0)$ &
\multicolumn{2} {|c|} {$(0,y,z)$} & \multicolumn {2} {|c|} {$(0,y,z)$} &
\multicolumn{2} {|c|} {$(0,y,z)$} & \multicolumn {3} {|c|} {$(x,y,z)$} & 
& &  \\ 
& $y$ & $z$ & $y$ & $y$ & $z$ & $y$ & $z$ & $y$ & $z$ & x & $y$ & $z$ &
& &  \\  \hline
\#4  & 0.023 & -0.446 & -0.119 & 0.040 & 0.169 &-0.032 & -0.015 & 0.025 & -0.314&
-0.024 & 0.005 & 0.048 & 9.2 & 0.47 & 2.5 \\ \hline
\#16 & 0.551 & -0.007 & -0.209 &-0.094 &-0.063 &-0.099 & -0.111 &-0.061 &  0.069&
 0.013 & 0.073  & -0.001 &21.3 & 0.31 & 1.7 \\ \hline
\#27 &-0.205 &  0.058 & -0.110 & 0.206 & 0.172 &-0.006 & -0.354 & 0.023 & -0.061&
 0.071 & 0.041  & 0.009 &30.4 & 0.26 & 3.3 \\ \hline
\#29 &-0.008 & -0.428&  0.019 & 0.010 &-0.265 & 0.090 &  0.050 &-0.053 &  0.086&
-0.020 &-0.040  &-0.178 &31.1 & 0.26 & 6.1 \\ \hline
\#34 &-0.164 &  0.066 & -0.286 & 0.073 & 0.049 & 0.109 &  0.163 & 0.103 &  0.077&
-0.073 & 0.188  &-0.068 &36.2 & 0.24 &66.1 \\ \hline
\#40 & 0.178 &  0.113 & -0.166 & 0.162 &-0.012 & 0.157 &  0.076 & 0.171 & -0.012&
 0.029&-0.237  & 0.007 &38.9 & 0.23 &18.7 \\ \hline
\#49 & 0.069 &  0.307 &  0.021 &-0.056 &-0.138 & 0.023 &  0.057 &-0.069 & -0.348&
-0.068 & 0.015  &-0.139 &44.4 & 0.22 & 0.5\\ \hline
\#53 &-0.014 & -0.022 & -0.037 &-0.043 & 0.165 & 0.143 &  0.182 &-0.112 & -0.080&
 0.212 & 0.066  & 0.155 &49.8 & 0.20 & 16.3 \\ \hline
\#64 & 0.037 &  0.041 & -0.030 & 0.264 & 0.110 & 0.015 &  0.047 &-0.396 &  0.042&
-0.038 &-0.027  &-0.010 &69.2 & 0.17 & 46.1 \\ \hline
\#70 & 0.011 &  -0.011 &  0.001 &-0.155 & 0.017 & 0.365 & -0.169 &-0.081 &  0.011&
-0.143 & -0.004   & 0.090 &93.3 & 0.15 & 23.0 \\ \hline
\#78 & 0.010 &  0.019 &  0.004 &-0.089 & 0.238 & 0.150 & -0.099 &-0.029 &  0.000&
 0.203 & 0.009   &-0.187 &103.0& 0.14 & 11.3\\ \hline
\end{tabular}
\caption{\label{b2u} Mass-weighted atomic displacements of B$_{2u}$ phonons which induce
a dipole moment along the ${\bf b}$ axis (normalized so that the sum of the
squares of the components equals unity for each mode).  (The acoustic B$_{2u}$
is not tabulated.)  Each component of the
mass-weighted displacement represents the atomic
displacement divided by the square root of the respective atomic mass.  The
mass-weighted displacements are given for the sites listed in column
\#2 of Table \ref{orbit}.  The displacements of the remaining atoms in
each Wyckoff orbit are fixed so that the mode transforms like B$_{2u}$, i. e.
like the $y$-component of a vector.  The calculated atomic charges, magnitude
of the rms displacement $Q_{\rm rms}$ (where $m_p$ is the proton mass),
the rms dipole moment, and the mode energy is also given.}
\end{table*}

\begin{figure}
\includegraphics[width=8cm]{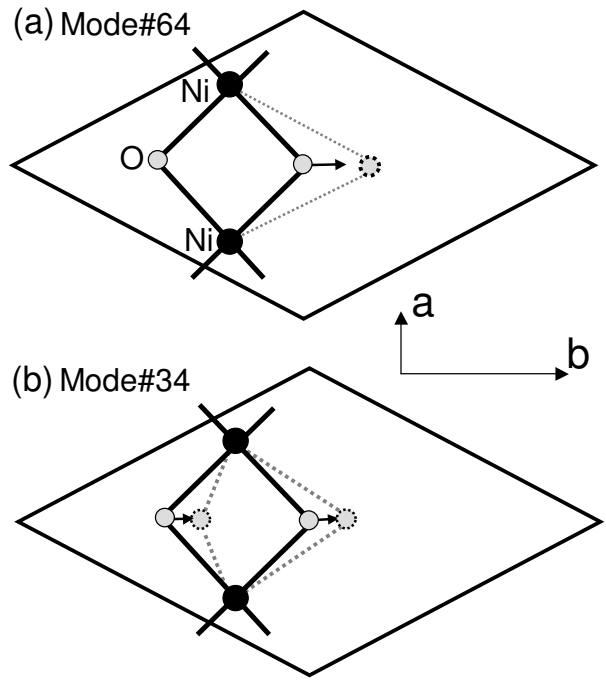}
\caption{\label{mode34_64} A schematic representation of the top view
of two particularly interesting B$_{2u}$ modes whose displacement vectors 
are given in Table~\ref{b2u}.}
\end{figure}

Figure~\ref{mode34_64} shows schematically
how the oxygen atoms move in these two particularly interesting phonons. 
For the low-energy mode at 36 meV, the two oxygen atoms
connecting the spine-spins move in the same direction. Therefore
while one of the Ni-O-Ni bond angle
decreases, the other Ni-O-Ni bond angle increases.
Hence at first order, we do not expect large changes in the Ni-O-Ni
superexchange due to this phonon. On the other hand, for the $E=70$
meV mode, only one oxygen (which is connected to the cross-tie Ni spin)
moves along the ${\bf b}$-axis. Hence, in
this case, only one of the Ni-O-Ni bond angles
changes from nearly $90^{\rm o}$ and
therefore we expect this phonon to have important effects on the
Ni-O-Ni superexchange interaction. Interestingly, the biggest disagreement
between the experimental data
and the calculated phonon energies happens for this phonon, which further
suggests that it may have strong spin-phonon coupling.

Finally, we discuss what kind of distortion is needed in order to
induce the experimentally observed dipole moment whose magnitude is
about $P_{\rm exp}=1.25\times10^{-4}$ C/m$^{2}$.  We note that
this induced dipole moment is much smaller than the calculated rms dipole
moment 
($P_{\rm rms}~ \sim 46-70 \times10^{-4}$ C/m$^{2}$) listed in Table~\ref{b2u}. 
This suggests that the local distortion should be at the order of
$P_{\rm exp}/P_{\rm rms}$, which suggests that atoms move less than
0.001 \AA. This is a quite small distortion and would be very difficult
to observe directly by neutron powder diffraction.

\section{Symmetry of the Spin Structure}

In Refs. \onlinecite{FERRO}, \onlinecite{EXPT}, \onlinecite{REV}, and
\onlinecite{THEORY} the application of
representation theory to the determination and characterization of
magnetic structures is discussed in detail. In Table \ref{CHARQ} we
give the character table for the irreps for an arbitrary wavevector
of the form $(q,0,0)$.  Because the irreps are one dimensional,
each spin basis function is an eigenvector of the symmetry operator
with the listed eigenvalue.

\begin{table}
\begin{tabular}{||cc|cccc||}\hline\hline
& &$1$ &  $2_x$ & $\tilde m_y$ & $\tilde m_z$ \\ \hline
& $\Gamma^1$ & 1 &  1 &  1 &  1 \\
& $\Gamma^2$ & 1 &  1 & $-1$ & $-1$ \\
& $\Gamma^3$ & 1 & $-1$ &  $1$ & $-1$ \\
& $\Gamma^4$ & 1 & $-1$ & $-1$ &  $1$ \\ \hline\hline
\end{tabular}
\caption{\label{CHARQ}Irreducible representations of
the group $G_{\bf v}$ for the incommensurate magnetic structure with
${\bf v}=(q,0,0)$. Here it is simplest to use the symmetry operations
$\tilde m_y$ and $\tilde m_z$, such that $\tilde m_y {\bf r}=
(x, \overline y + \oh , z + \oh )$ and $\tilde m_z {\bf r} =
(x , y + \oh , \overline z + \oh )$.}
\end{table}

In these references it is shown that
the HTI phase is described by a set of five complex amplitudes
associated with the irrep $\Gamma_4$.  Here we call these
$a_{s,x}$, $ia_{s,y}$, and $a_{s,z}$ to describe
the orientation of the spine spins and $a_{c,y}$ and $a_{c,z}$ to
describe the orientation of the cross-tie spins.  When the LTI phase is
entered additional variables associated with the irrep $\Gamma_2$
become nonzero.  These LTI variables are here denoted
$ib_{s,x}$, $b_{s,y}$, and $ib_{s,z}$ to describe
the orientation of the spine spins and $b_{c,z}$ to describe the
orientation of the cross tie spins.  Because the crystal
is centrosymmetric, it is shown\cite{EXPT,REV,THEORY}
that within a given representation all these complex 
structural parameters (the $a$'s and $b$'s) 
can be written in terms of a real amplitude times a complex
phase factor which is the same for all variables of the same
irrep, $\Gamma_4$ or $\Gamma_2$, in the sense that
\begin{eqnarray}
a_{t,\alpha} &=& \pm |a_{t,\alpha}| e^{i \phi_{HTI}} \ , \ \ \
b_{t,\alpha} = \pm |b_{t,\alpha}| e^{i \phi_{LTI}} \ ,
\label{PHASEEQ} \end{eqnarray}
where $t$ denotes either spine or cross-tie.  It is further
expected\cite{EXPT,THEORY} that
$\phi_{\rm HTI} - \phi_{\rm LTI}$ is $\pm \pi /2$.
Thus in these two phases we may use the results of Table VIII in
Ref. \onlinecite{EXPT} to write the spin components of
the six Ni ions in the unit cell as
\begin{eqnarray}
S_x^{(1)}({\bf R}_1) &=& (a_{s,x}+ib_{s,x}) e^{i{\bf q} \cdot {\bf R}_1 }
\nonumber \\ && \ + (a_{s,x}^* -ib_{s,x}^*) e^{-i{\bf q}\cdot {\bf R}_1 } 
\label{SPINS1} \\
S_y^{(1)}({\bf R}_1) &=& (ia_{s,y}+b_{s,y}) e^{i{\bf q} \cdot {\bf R}_1 }
\nonumber \\ && \ + (- ia_{s,y}^*+b_{s,y}^*) e^{-i{\bf q}\cdot{\bf R}_1 } \\
S_z^{(1)}({\bf R}_1) &=& (a_{s,z}+ib_{s,z}) e^{i{\bf q} \cdot {\bf R}_1 }
\nonumber \\ && \ + (a_{s,z}^*-ib_{s,z}^*) e^{-i{\bf q} \cdot {\bf R}_1 } \\
S_x^{(2)}({\bf R}_2) &=& (-a_{s,x}+ib_{s,x}) e^{i{\bf q} \cdot{\bf R}_2 }
\nonumber \\ && \ + (- a_{s,x}^* -ib_{s,x}^*)e^{-i{\bf q}\cdot{\bf R}_2 } \\
S_y^{(2)}({\bf R}_2) &=& (ia_{s,y}-b_{s,y})e^{i{\bf q}   \cdot{\bf R}_2 }
\nonumber \\ && \ + (-i a_{s,y}^*-b_{s,y}^*)e^{-i{\bf q} \cdot{\bf R}_2 } \\
S_z^{(2)}({\bf R}_2) &=& (a_{s,z}-ib_{s,z}) e^{i{\bf q}  \cdot{\bf R}_2 }
\nonumber \\ && \ + (a_{s,z}^*+ib_{s,z}^*) e^{-i{\bf q}  \cdot{\bf R}_2 } \\
S_x^{(3)}({\bf R}_3) &=& (a_{s,x}-ib_{s,x}) e^{i{\bf q}  \cdot{\bf R}_3 }
\nonumber \\ && \ + (a_{s,x}^*+ib_{s,x}^*) e^{-i{\bf q}  \cdot{\bf R}_3 } \\
S_y^{(3)}({\bf R}_3) &=& (-ia_{s,y}+b_{s,y}) e^{i{\bf q} \cdot{\bf R}_3 }
\nonumber \\ && \ +(i a_{s,y}^*+b_{s,y}^*) e^{-i{\bf q}  \cdot{\bf R}_3 } \\
S_z^{(3)}({\bf R}_3) &=& (a_{s,z}-ib_{s,z}) e^{i{\bf q}  \cdot{\bf R}_3 }
\nonumber \\ && \ + (a_{s,z}^*+ib_{s,z}^*) e^{-i{\bf q}  \cdot{\bf R}_3 } \\
S_x^{(4)}({\bf R}_4) &=& (-a_{s,x}-ib_{s,x}) e^{i{\bf q} \cdot{\bf R}_4 }
\nonumber \\ && \ + (-a_{s,x}^*+ib_{s,x}^*) e^{-i{\bf q} \cdot{\bf R}_4 } \\
S_y^{(4)}({\bf R}_4) &=& (-ia_{s,y}-b_{s,y}) e^{i{\bf q} \cdot{\bf R}_4 }
\nonumber \\ && \ +(ia_{s,y}^*-b_{s,y}^*) e^{-i{\bf q}   \cdot{\bf R}_4 } \\
S_z^{(4)}({\bf R}_4) &=& (a_{s,z}+ib_{s,z}) e^{i{\bf q}  \cdot{\bf R}_4 }
\nonumber \\ && \ + (a_{s,z}^*-ib_{s,z}^*) e^{-i{\bf q}  \cdot{\bf R}_4 } \\
S_x^{(5)}({\bf R}_5) &=& b_{c,x} e^{i{\bf q} \cdot {\bf R}_5}
+ b_{c,x}^* e^{-i{\bf q} \cdot {\bf R}_5} \\
S_y^{(5)}({\bf R}_5) &=& a_{c,y} e^{i{\bf q} \cdot {\bf R}_5}
+ a_{c,y}^* e^{-i{\bf q} \cdot {\bf R}_5} \\
S_z^{(5)}({\bf R}_5) &=& a_{c,z} e^{i{\bf q} \cdot {\bf R}_5}
+ a_{c,z}^* e^{-i{\bf q} \cdot {\bf R}_5} \\
S_x^{(6)}({\bf R}_6) &=& -b_{c,x} e^{i{\bf q} \cdot {\bf R}_6}
-b_{c,x}^* e^{-i{\bf q} \cdot {\bf R}_6} \\
S_y^{(6)}({\bf R}_6) &=& -a_{c,y} e^{i{\bf q} \cdot {\bf R}_6}
- a_{c,y}^* e^{-i{\bf q} \cdot {\bf R}_6} \\
S_z^{(6)}({\bf R}_6) &=& a_{c,z} e^{i{\bf q} \cdot {\bf R}_6} 
+ a_{c,z}^* e^{-i{\bf q} \cdot {\bf R}_6} \ .
\label{SPINS2} \end{eqnarray}
Here the superscripts on 1, 2, 3, 4 on ${\bf S}$ label the spine
sites and 5 and 6 label the cross-tie sites c and c$^\prime$,
respectively, (as in Fig. \ref{SPSP}, below)
and ${\bf R}_n$ is the position of a spin $n$.  Also
${\bf q}=q\hat i$ with $q \approx 0.28(2 \pi /a)$.\cite{LawesKenzelmann,EXPT}

\section{Relations for the Strain Derivative of the Exchange Tensor}

In this section we obtain explicit forms for the most important spin-phonon
coupling matrices. For this purpose we start by introducing notation
for the principal exchange interactions.  We write the interactions between
spins on sites $i$ and $j$ as
\begin{eqnarray}
{\cal H} (i,j) &=& \sum_{\alpha \beta} X_{\alpha \beta} (i,j)
S_\alpha (i) S_\beta (j) \ ,
\end{eqnarray}
where $X_{\alpha \beta} (i,j) = X_{\beta \alpha}(j,i)$, of course.
For nearest neighbor (nn) interactions between spine spins we set 
$X_{\alpha \beta} (i,j)=U_{\alpha \beta} (i,j)$, for next nearest neighbor
(nnn) interactions between spine spins we set
$X_{\alpha \beta} (i,j)=V_{\alpha \beta} (i,j)$, and for nn interactions
between spine and cross-tie spins we set
$X_{\alpha \beta} (i,j)=W_{\alpha \beta} (i,j)$.
We may further decompose the exchange tensor into its symmetric
and antisymmetric parts.  For example, for nn spine-spine interactions
we write (omitting the site labels $i$ and $j$)
\begin{eqnarray}
{\bf U} &=& \left[ \begin{array} {c c c}
J_{xx} & J_{xy} + D_z & J_{xz} - D_y \\
J_{xy} - D_z & J_{yy} & J_{yz}  + D_x \\
J_{xz} + D_y & J_{yz} - D_x & J_{zz} \\
\end{array} \right] \ ,
\end{eqnarray}
where ${\bf D}$ is the Dzyaloshinskii-Moriya (DM) vector.\cite{DM1,DM2}
Similar decompositions will be made for ${\bf V}$ and ${\bf W}$ in terms
of symmetric tensors ${\bf K}$ and ${\bf L}$, respectively, and the
DM vectors ${\bf E}$ and ${\bf F}$, respectively.

Now we consider the gradient expansion of these exchange tensors.
From the above development we write the displacement ${\bf u}$
of the $\tau$th atom in the unit cell in terms of the amplitude
$Q_{\gamma_p}$ of the GD labeled $\gamma_p$ as
\begin{eqnarray}
u_\alpha ({\bf R}, \tau ; \gamma_p) &=&
Q_{\gamma_p} a(\gamma_p; \tau, \alpha ) \ ,
\end{eqnarray}
where $a(\gamma_p; \tau, \alpha )$ is the $\alpha$-component of the displacement
of atom $\tau$ in unit cell at ${\bf R}$ for the GD
labeled $\gamma_p$ (which are shown in Figs. \ref{GENDIS},
\ref{GENDIS2}, and \ref{GENDIS3}).  If $Z$
represents a component of an exchange tensor, then we write
\begin{eqnarray}
{\partial Z \over \partial Q_{\gamma_p}} &=& \sum_{\alpha {\bf R} \tau}
{\partial Z \over \partial u_\alpha ({\bf R}, \tau; \gamma_p)}
a(\gamma_p; \tau, \alpha ) \ .
\end{eqnarray}
The aim of the present paper is to determine which such derivatives are
required to completely determine the trilinear spin-phonon coupling.
The actual calculation of these coefficients will be undertaken separately.
Thus, for each GD $\gamma_p$, we consider the interaction
\begin{eqnarray}
{\cal H}_{\gamma_p} &=& Q_{\gamma_p} \sum_{\alpha \beta} \sum_{ij}
{\partial X_{\alpha \beta} (i,j) \over \partial Q_{\gamma_p} }
S_\alpha (i) S_\beta (j) \ .
\label{FEINT} \end{eqnarray}
Our objective is to express the results for the spontaneous polarization
due to the trilinear coupling in terms of the parameters    
$\partial X_{\alpha \beta} (i,j) / \partial Q_{\gamma_p}$.

\begin{figure}
\includegraphics[height=5.5cm]{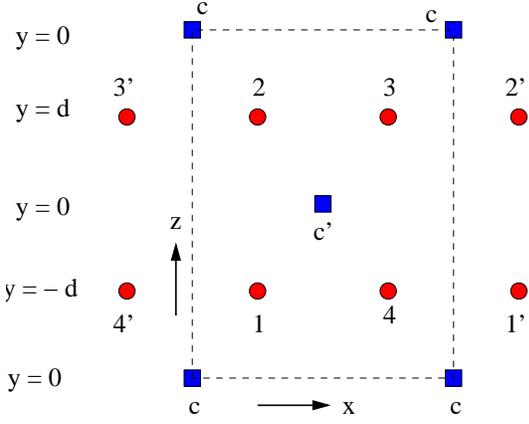}
\caption{\label{SPSP} Diagram of an ${\bf a}-{\bf c}$ plane
used to specify nearest-neighbor and next-nearest-neighbor interactions
along a single spine. Circles are spine sites and square are cross-tie
sites and $d=0.13b$ (see Table \protect{\ref{orbit}}).
The dashed rectangle indicates the unit cell.  Interactions in
other ${\bf a}-{\bf c}$ planes are obtained by using translation symmetry.}
\end{figure}

Here we analyze the gradients of the nn interactions between spine spins.
(Similar analyses of next-nearest neighbor spine interactions
and of nn spine-cross tie interactions are given in Appendices.)
We introduce the coupling between spine sites \#1 and
\#4 in Fig.  \ref{SPSP}:
\begin{eqnarray}
{\partial U_{\alpha \beta} (1,4) \over \partial Q_{\gamma_p}} &\equiv &
U_{\alpha \beta}^{\gamma_p} \ .
\end{eqnarray}
Clearly ${\bf U}^{\gamma_p}$ has to be invariant under the symmetry operations
of the crystal.  But $m_x$ takes the bond in question into itself
(and interchanges indices).
This indicates that these interaction matrices must satisfy
\begin{eqnarray}
\sigmav_x {\bf U}^{x_p} \sigmav_x  &=& - \tilde {\bf U}^{x_p} \nonumber \\
\sigmav_x {\bf U}^{y_p} \sigmav_x  &=& \tilde {\bf U}^{y_p} \nonumber \\
\sigmav_x {\bf U}^{z_p} \sigmav_x  &=& \tilde {\bf U}^{z_p} \ ,
\label{SIGEQ} \end{eqnarray}
where tilde indicates transpose, 
\begin{eqnarray}
\sigmav_x &=& \left[ \begin{array} {c c c} -1 & 0 & 0 \\ 0 &1 & 0 
\\ 0 & 0 & 1 \\ \end{array} \right] \ , \nonumber
\end{eqnarray}
and later
\begin{eqnarray}
\sigmav_y &=& \left[ \begin{array} {c c c} 1 & 0 & 0 \\ 0 & -1
& 0 \\ 0 & 0 & 1 \\
\end{array} \right] \ , \nonumber
\end{eqnarray}
\begin{eqnarray}
\sigmav_z &=& \left[ \begin{array} {c c c} 1 & 0 & 0 \\ 0 & 1 & 0
\\ 0 & 0 & -1 \\ \end{array} \right] \ .
\end{eqnarray}
Here we used the fact that the GD's were constructed to have known symmetry:
\begin{eqnarray}
m_\alpha Q_{\beta_p} = (1 - 2 \delta_{\alpha ,\beta}) Q_{\beta_p} 
\label{SIGEQ1}
\end{eqnarray}
and
\begin{eqnarray}
2_\alpha Q_{\beta_p} &=& (-1 + 2 \delta_{\alpha , \beta}) Q_{\beta_p} \ ,
\label{SIGEQ2}
\end{eqnarray}
where $\delta_{a,b}$ is unity if $a=b$ and is zero otherwise.

In view of Eq. (\ref{SIGEQ}), we have
\begin{eqnarray}
{\bf U}^{x_p} &=& \left[ \begin{array} {c c c}
0 & J_{xy}^x & J_{xz}^x \\ J_{xy}^x & 0 & D_x^x \\ J_{xz}^x & -D_x^x & 0 \\ 
\end{array} \right] \ , \nonumber  \\
{\bf U}^{y_p} &=& \left[ \begin{array} {c c c}
J_{xx}^y & D_z^y & -D_y^y \\ -D_z^y & J_{yy}^y &
J_{yz}^y \\ D_y^y & J_{yz}^y & J_{zz}^y \\ \end{array} \right] \ ,
\nonumber \\ {\bf U}^{z_p} &=& \left[ \begin{array} {c c c}
J_{xx}^z & D_z^z & -D_y^z \\ -D_z^z & J_{yy}^z & J_{yz}^z \\
D_y^z & J_{yz}^z & J_{zz}^z \\ \end{array} \right] \ ,
\end{eqnarray}
where the index $p$ on the superscripts of $J$ and $D$ are left implicit
and $J_{\alpha \beta}^{\gamma_p}$ and $D_\alpha^{\gamma_p}$
(and similarly later for superscripts on ${\bf K}$, ${\bf L}$, ${\bf E}$,
and ${\bf F}$) are defined to be
\begin{eqnarray}
J^{\gamma_p}_{\alpha \beta}  &\equiv &
{\partial J_{\alpha \beta} (1,4) \over \partial Q_{\gamma_p}} \ , \ \ \
D^{\gamma_p}_\alpha \equiv
{\partial D_\alpha (1,4) \over \partial Q_{\gamma_p}} \ .
\end{eqnarray}
Then we obtain the \#2-\#3 interaction from the above by $2_x$, a two-fold
rotation about the $x$-axis, so that
\begin{eqnarray}
{\partial U_{\alpha \beta} (2,3) \over \partial Q_{x_p}} &= &
\sigmav_y \sigmav_z  {\bf U}^{x_p} \sigmav_y \sigmav_z \nonumber \\ &=&
\left[ \begin{array} {c c c} 0 & -J_{xy}^x & - J_{xz}^x \\
-J_{xy}^x & 0 & D_x^x \\
-J_{xz}^x & - D_x^x & 0 \\ \end{array} \right] \ ,
\end{eqnarray}
\begin{eqnarray}
{\partial U_{\alpha \beta} (2,3) \over \partial Q_{y_p}} &= &
- \sigmav_y \sigmav_z {\bf U}^{y_p} \sigmav_y \sigmav_z \nonumber \\ &=&
\left[ \begin{array} {c c c}
-J_{xx}^y & D_z^y & -D_y^y \\ -D_z^y & -J_{yy}^y &
-J_{yz}^y \\ D_y^y & -J_{yz}^y & -J_{zz}^y \\ \end{array} \right] \ ,
\end{eqnarray}
\begin{eqnarray}
{\partial U_{\alpha \beta} (2,3) \over \partial Q_{z_p}} &= &
- \sigmav_y \sigmav_z {\bf U}^{z_p} \sigmav_y \sigmav_z  \nonumber \\ &=&
\left[ \begin{array} {c c c}
-J_{xx}^z & D_z^z & -D_y^z \\ -D_z^z & -J_{yy}^z & -J_{yz}^z \\
D_y^z & -J_{yz}^z & -J_{zz} \\ \end{array} \right] \ .
\end{eqnarray}
where we used Eq. (\ref{SIGEQ2}).

We obtain the \#4-\#1' interactions by applying the glide operation $m_y$  
to the \#2-\#3 interaction, so that
\begin{eqnarray}
{\partial U_{\alpha \beta} (4,1') \over \partial Q_{x_p}} &= &
\sigmav_y {\partial {\bf U}(2,3) \over \partial Q_{x_p}} \sigmav_y
\nonumber \\ &=&
\left[ \begin{array} {c c c} 0 & J_{xy}^x & - J_{xz}^x \\ 
J_{xy}^x & 0 & -D_x^x \\ -J_{xz}^x & D_x^x & 0 \\ \end{array} \right] \ ,
\end{eqnarray}
\begin{eqnarray}
{\partial U_{\alpha \beta} (4,1') \over \partial Q_{y_p}} &= &
- \sigmav_y {\partial {\bf U}(2,3) \over \partial Q_{y_p}} \sigmav_y
\nonumber \\ &=&
\left[ \begin{array} {c c c}
J_{xx}^y & D_z^y & D_y^y \\ -D_z^y & J_{yy}^y &
-J_{yz}^y \\ -D_y^y & -J_{yz}^y & J_{zz}^y \\ \end{array} \right] \ ,
\end{eqnarray}
\begin{eqnarray}
{\partial U_{\alpha \beta} (4,1') \over \partial Q_{z_p}} &= &
\sigmav_y {\partial {\bf U}(2,3) \over \partial Q_{z_p}} \sigmav_y
\nonumber \\ &=&
\left[ \begin{array} {c c c}
-J_{xx}^z & -D_z^z & -D_y^z \\ D_z^z & -J_{yy}^z & J_{yz}^z \\
D_y^z & J_{yz}^z & -J_{zz}^z \\ \end{array} \right]\ ,
\end{eqnarray}
and finally we get the \#3-\#2' interaction by applying a two-fold
rotation about the $x$-axis to the \#4-\#1' interaction to get
\begin{eqnarray}
{\partial U_{\alpha \beta} (3,2') \over \partial Q_{x_p}} &= &
\sigmav_y \sigmav_z {\partial {\bf U} (4,1') \over \partial Q_{x_p}}
\sigmav_y \sigmav_z \nonumber \\ &=&  
\left[ \begin{array} {c c c} 0 & -J_{xy}^x & J_{xz}^x \\ -J_{xy}^x & 0 & -D_x^x \\
J_{xz}^x & D_x^x & 0 \\ \end{array} \right] \ ,
\end{eqnarray}
\begin{eqnarray}
{\partial U_{\alpha \beta} (3,2') \over \partial Q_{y_p}} &= &
- \sigmav_y \sigmav_z {\partial {\bf U} (4,1') \over \partial Q_{y_p}}
\sigmav_y \sigmav_z \nonumber \\ &=&  
\left[ \begin{array}{c c c}
-J_{xx}^y & D_z^y & D_y^y \\ -D_z^y & -J_{yy}^y &
J_{yz}^y \\ -D_y^y & J_{yz}^y & -J_{zz}^y \\ \end{array} \right]\ ,
\end{eqnarray}
and
\begin{eqnarray}
{\partial U_{\alpha \beta} (3,2') \over \partial Q_{z_p}} &= &
- \sigmav_y \sigmav_z {\partial {\bf U} (4,1') \over \partial Q_{z_p}}
\sigmav_y \sigmav_z \nonumber \\ &=&  
\left[ \begin{array}{c c c}
J_{xx}^z & -D_z^z & -D_y^z \\ D_z^z & J_{yy}^z &
-J_{yz}^z \\ D_y^z & -J_{yz}^z & J_{zz}^z \\ \end{array} \right]\ .
\end{eqnarray}

\section{Mean Field Spin-Phonon Hamiltonian}

\subsection{Mean Field Results}

Here we treat the nn spine-spine interactions in detail. Analogous
calculations for the nnn spine-spine and nn spine-cross tie
interactions are treated in an Appendix.  
We evaluate the spin-phonon Hamiltonian ${\cal H}_{\gamma_p}$
of Eq. (\ref{FEINT}) at the mean-field level.  In other words,
for the spin operators we simply substitute their average
values as given in Eqs. (\ref{SPINS1})-(\ref{SPINS2}).
One sees that ${\cal H}_{x_p}$ for nn spine-spine interactions, for
instance, will consist of
contributions proportional to $J_{xy}^{x_p}$, to
$J_{xz}^{x_p}$, and to $D_x^{x_p}$.  To illustrate the
calculation we will explicitly evaluate the first of these,
which we denote ${\cal H}_{x_p}(J_{xy})$:
\begin{eqnarray}
{\cal H}_{x_p}(J_{xy})&=& Q_{x_p} J_{xy}^{x_p}
\sum_{uc} \left[ S_x^{(1)} S_y^{(4)} + S_y^{(1)} S_x^{(4)} \right. 
\nonumber \\ && \ - S_x^{(2)} S_y^{(3)} - S_x^{(3)} S_y^{(2)}
+ S_x^{(4)} S_y^{(1')} \nonumber \\ && \ \left.
+ S_x^{(1')} S_y^{(4)} - S_x^{(3)} S_y^{(2')} 
- S_x^{(2')} S_y^{(3)} \right] \ ,
\end{eqnarray}
where, since we included all interactions within a single unit
cell, the sum is over all $N_{\rm u c}$ unit cells. (In this
summation only terms involving both ${\bf q}$ and $-{\bf q}$
survive.)  Thus
\begin{eqnarray}
{\cal H}_{x_p}(J_{xy})&=&
2 N_{\rm u c} Q_{x_p} J_{xy}^{x_p} e^{-iqa/2} \nonumber \\
&& \ \times \bigl[
(a_{s,x} + i b_{s,x})(ia_{s,y}^* - b_{s,y}^*)
\nonumber \\ && \ +
(ia_{s,y} + b_{s,y})(-a_{s,x}^* + ib_{s,x}^*)
\nonumber \\ && \
- (-a_{s,x} + i b_{s,x})(ia_{s,y}^* + b_{s,y}^*)
\nonumber \\ && \
- (ia_{s,y} - b_{s,y})(a_{s,x}^* + ib_{s,x}^*) \bigr] + {\rm c. \ c.}
\nonumber \\ &=&
16 N_{\rm u c} Q_{x_p} J_{xy}^{x_p} \cos (qa/2) \nonumber \\
&& \ \times \Im
[a_{s,x}^* a_{s,y} + b_{s,x} b_{s,y}^*] \ .
\end{eqnarray}
The other terms proportional to $Q_{x_p}$ are
\begin{eqnarray}
{\cal H}_{x_p}(J_{xz})&=& 16 N_{\rm u c} Q_{x_p} J_{xz}^{x_p} \sin (qa/2)
\nonumber \\ && \ \times 
\Im [a_{s,x} a_{s,z}^* + b_{s,x} b_{s,z}^*]  \ .
\end{eqnarray}
\begin{eqnarray}
{\cal H}_{x_p}(D_x)&=& 16 N_{\rm u c} Q_{x_p} D_x^{x_p} \cos (qa/2)
\nonumber \\ && \ \times
\Im [-a_{s,y} a_{s,z}^* + b_{s,y} b_{s,z}^*] \ .
\end{eqnarray}
In view of Eq. (\ref{PHASEEQ}) all these terms involving $Q_{x_p}$
vanish, as was found from the phenomenological formulation.
Similarly, all the terms involving $Q_{z_p}$ also vanish, again
in conformity with the phenomenological argument.

We are thus only left with terms involving $Q_{y_p}$. In Appendix
C we find that the strain dependence of the nn spine interactions give
\begin{eqnarray}
{\cal H}_{y_p} &=& 16 N_{uc} Q_{y_p} \sum_{\mu , \nu = x,y,z}
\Lambda_{\mu \nu}^{(nn)} \  \Im \left[ a_{s,\mu}^* b_{s,\nu} \right] \ ,
\label{YPEQ} \end{eqnarray}
where
\begin{eqnarray}
\Lambdav^{(nn)} &=& \left[ \begin{array} {c c c}
J_{xx}^{y_p} c & D_z^{y_p} s & D_y^{y_p}c \\
- D_z^{y_p} s & -J_{yy}^{y_p} c &
- J_{yz}^{y_p} s \\ D_y^{y_p} c &
J_{yz}^{y_p} s& - J_{zz}^{y_p} c \\
\end{array} \right] \ ,
\label{G1EQ} \end{eqnarray}
where $c\equiv \cos (qa/2)$ and $s \equiv \sin (qa/2)$.
Using the results of Appendix C we find that the nnn
interactions give a result of the form of Eq. (\ref{YPEQ})
but with
\begin{eqnarray}
\Lambdav^{(nnn)} &=& \left[ \begin{array} {c c c}
- K_{xx}^{y_p} c' & - E_z^{y_p} s' & - K_{xz}^{y_p}c' \\
E_z^{y_p} s' & K_{yy}^{y_p} c' &
 -E_x^{y_p} s' \\ -K_{xz}^{y_p} c' &
E_x^{y_p} s' & -K_{zz}^{y_p} c' \\
\end{array} \right] \ ,
\label{YPEQ2} \end{eqnarray}
where $c' \equiv \cos (qa)$ and $s' \equiv \sin (qa)$.
Using the results of Appendix C we have the results for
the spine-cross tie exchange gradients:
\begin{eqnarray}
V_{y_p}&=& 16 N_{\rm u c} Q_{y_p} \left(
\sum_{\alpha=y,z} \sum_\beta \Lambda_{\alpha \beta}^{sx}
\Im \left[ a_{c \alpha} b_{s \beta}^* \right] \right.
\nonumber \\ && \ \left.
+ \sum_{\alpha=x} \sum_\beta \Lambda_{\alpha \beta}^{sx}
\Im \left[ b_{c \alpha} a_{s, \beta}^* \right] \right) \ ,
\label{YPEQ3} \end{eqnarray}
where $\Lambdav^{sx}$ is
\begin{eqnarray}
\left[ \begin{array} {c c c}
L_{xx}^{y_p} {s}^{\prime \prime} & \left[ L_{xy}^{y_p} + F_z^{y_p} 
\right] c'' &
\left[ L_{xz}^{y_p}-F_y^{y_p} \right] {s}^{\prime \prime} \\
\left[ L_{xy}^{y_p} - F_z^{y_p} \right] {c}^{\prime \prime} &
L_{yy}^{y_p} {s}^{\prime \prime} &
\left[ L_{yz}^{y_p} + F_x^{y_p} \right] {c}'' \\
\left[ L_{xz}^{y_p}+F_y^{y_p} \right] {c}'' &
\left[ L_{yz}^{y_p}-F_x^{y_p} \right] {s}'' & L_{zz}^{y_p} {c}'' \\
\end{array} \right] \ ,
\label{G3EQ} \end{eqnarray}
where $c'' \equiv \cos (qa/4)$ and $s'' \equiv \sin (qa/4)$.
These results agree with the phenomenological model, in that
$V_{y_p}$ 
is only nonzero when both the ``$a$" and the ``$b$" irreps are
simultaneously present and they can not have the same phase
(lest $a^*b$ be real).

\subsection{Summary} \label{RESULTS}

Here we show how the above results lead to an evaluation of the
spontaneous polarization. If we combine the results of Eqs. 
(\ref{YPEQ}), (\ref{YPEQ2}), and (\ref{YPEQ3}), we may write
\begin{eqnarray}
V_{y_p} = N_{\rm uc} Q_{y_p} \lambda_{y_p} \ , \ \ \ 
p=1,12 \ .
\end{eqnarray}
But the GD's are related to the normal modes via
\begin{eqnarray}
Q_{y_p} &=& \sum_{n=1,11} o_{y_p}^{(n)} M_p^{-1/2} Q_ n \ ,
\end{eqnarray}
where the orthogonal transformation ${\bf o}$ is essentially that
given in Table \ref{b2u} apart from normalization factors due
to replacing a Wyckoff orbit by one of its atoms.  Note that the GD
$y_p$ is associated with atoms all having the same mass $M_p$.  The
magnetoelectric trilinear interaction in terms of normal mode operators is
thus
\begin{eqnarray}
V &=& N_{uc} \sum_{n,p} \lambda_{y_p} o_{y_p}^{(n)} M_p^{-1/2} Q_n \ ,
\end{eqnarray}
Since the unperturbed elastic energy is ${\cal H}_0 = (N_{\rm uc}/2)
\sum_n \omega_n^2 Q_n^2$, we see that the trilinear interaction
induces the displacements
\begin{eqnarray}
\langle Q_n \rangle = \sum_p 
\lambda_{y_p} o_{y_p}^{(n)} M_p^{-1/2} \omega_n^{-2} \ ,
\end{eqnarray}
from which we obtain the induced spontaneous polarization as
\begin{eqnarray}
P_y &=& {1 \over \Omega_{\rm uc}} \sum_{n, p}
q_p o_{y_p}^{(n)} \langle Q_n \rangle M_p^{-1/2} \ ,
\label{PEQ} \end{eqnarray}
where the atomic charges $q_p$ are given in Table \ref{b2u}. 
The only ingredients we do not have for this evaluation are
the various gradients of the exchange tensors.

\section{Toy Models} \label{TOYSEC}
 
In this section we will consider two toy models.  In the first one, we consider a single spine
with one Ni per unit cell, but with two oxygen atoms symmetrically placed on either side
of the spine. In this version, all atoms lie in a single plane.  In the second version,
the size of the unit cell is doubled.  In one plaquette the oxygen atoms are both displaced
equally perpendicularly to the original atomic plane and in the next plaquette the oxygens
are oppositely displaced.  In this version one of the mirror planes becomes a glide plane.
These models illustrate the simplifications which arise when the system has higher
symmetry than the buckled kagom\'{e} lattice of NVO.

\subsection{Unit cell with two atoms}

\begin{figure}
\includegraphics[width=8cm]{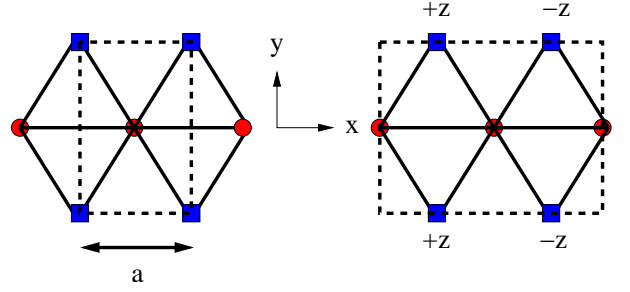}
\caption{\label{TOY} The two toy models.  In each case the unit cell
is bounded by a dashed line.  The filled circles are Ni sites
and the squares are O sites.}
\end{figure}

In this section we consider the toy model shown in the left panel of Fig. \ref{TOY}.
We first analyze the symmetry of the strain derivatives of the exchange
tensor.  By translational symmetry all nearest neighbor interactions are
equivalent.  So we set
\begin{eqnarray}
&& {\partial J_{\alpha \beta} (n,n+1) \over \partial Q_{\gamma_p}} \equiv
{\bf J}_{\alpha \beta}^{\gamma_p} \nonumber \\ && \ \ =
\left[ \begin{array} {c c c} J_{xx}^\gamma & J_{xy}^\gamma + D_z^\gamma & 
J_{xz}^\gamma - D_y^\gamma \\ J_{xy}^\gamma - D_z^\gamma &
J_{yy}^\gamma & J_{yz}^\gamma + D_x^\gamma \\ J_{xz}^\gamma
+ D_y^\gamma & J_{yz}^\gamma - D_x^\gamma & J_{zz}^\gamma \\
\end{array} \right] \ ,
\end{eqnarray}
where $J_{\alpha \beta}^\gamma \equiv \partial J_{\alpha \beta} /
\partial Q_{\gamma_p}$, $D_\alpha^\gamma \equiv \partial D_\alpha
/ \partial Q_{\gamma_p}$,  and the index $p$ is left implicit. The
Hamiltonian is invariant under mirror reflections with respect to
each coordinate axis.  Taking account of the symmetry of the displacement
coordinate and the fact that $m_x$
interchanges indices of the exchange tensor, we have that
\begin{eqnarray}
\sigmav_x {\bf J}^\gamma \sigmav_x &=&
(1 - 2 \delta_{x, \gamma}) \tilde {\bf J}^\gamma \nonumber \\  
\sigmav_y {\bf J}^\gamma \sigmav_y &=&
(1 - 2 \delta_{y, \gamma}) {\bf J}^\gamma \nonumber \\  
\sigmav_z {\bf J}^\gamma \sigmav_z &=&
(1 - 2 \delta_{z, \gamma}) {\bf J}^\gamma \ .
\label{SYMEQS} \end{eqnarray}

\noindent
As a result of this symmetry we have that ${\bf J}^x = 0$,
\begin{eqnarray}
{\bf J}^y&=& \left[ \begin{array} {c c c}
0 & D_z^y & 0 \\ -D_z^y & 0 & 0 \\ 0 & 0 & 0 \\ \end{array}
\right] \ , \nonumber \\ 
{\bf J}^z&=& \left[ \begin{array} {c c c}
0 & 0 & -D_y^z \\ 0 & 0 & 0 \\ D_y^z & 0 & 0 \\ \end{array}
\right] \ .
\end{eqnarray}
We construct the trilinear spin-phonon interaction, $V$, as
\begin{eqnarray}
&& V = \sum_p \left( Q_{y_p} D_z^{y_p} C_y + Q_{z_p} D_y^{z_p} C_z
\right) \ ,
\end{eqnarray}
where
\begin{eqnarray}
C_\alpha &=& \sum_n [ S_x(n) S_\alpha (n+1) - S_\alpha (n) S_x(n+1) ] .
\end{eqnarray}
Now we replace the spins by their equilibrium values.  We note that
each spin component belongs to a separate representation and we write
\begin{eqnarray}
S_x(n) = S_x(q) e^{inqa} + S_x(q)^* e^{-inqa} \ , \nonumber \\
S_y(n) = S_y(q) e^{inqa} + S_y(q)^* e^{-inqa} \ , \nonumber \\
S_z(n) = S_z(q) e^{inqa} + S_z(q)^* e^{-inqa} \ .
\end{eqnarray}
Then, keeping only those terms which survive the sum over $n$ we have
\begin{eqnarray}
V&=& 4N \sin (qa) \sum_p \left[ Q_{y_p} D_z^{y_p} r_x(q) r_y(q) \sin(\phi_x
-\phi_y) \right. \nonumber \\ && \ \left.
 + Q_{z_p} D_y^{z_p} r_z(q) r_x(q) \sin(\phi_x - \phi_z) \right] \ ,  
\end{eqnarray}
where $N$ is the total number of Ni spins and
we set $S_\alpha (q) = r_\alpha (q) e^{i \phi_\alpha}$, where
$r_\alpha (q)$ is real.  As found before\cite{FERRO,EXPT,REV} this
interaction is only nonzero when a) two different representations
are condensed and b) the two representation have different phases $\phi$.
In view of our previous results, it is clear that the appearance of
only strain derivatives of the DM vector is an artifact of the
rather high symmetry of this coplanar model.

\subsection{Unit cell with four atoms} 

Now we consider the noncoplanar toy model shown in the right panel of
Fig. \ref{TOY}.  The first two symmetry relations of Eq. (\ref{SYMEQS})
remain valid, but the third one now results from the glide plane which
involves a translation along the chain.  If ${\bf J}^\alpha_-$ denotes
the strain derivative of the exchange tensor for coupling sites $2n$ and
$2n+1$, and ${\bf J}_+^\alpha$ that for sites $2n+1$ and $2n+2$, then we have
\begin{eqnarray}
{\bf J}_\pm^x &=& \left[ \begin{array} {c c c} 0 & 0 & \pm J_{xz}^x \\
0 & 0 & 0 \\ \pm J_{xz}^x & 0 & 0 \\ \end{array} \right] \ ,
\end{eqnarray}
\begin{eqnarray}
{\bf J}_\pm^y &=& \left[ \begin{array} {c c c} 0 & D_z^y & 0 \\
-D_z^y & 0 & \pm J_{yz}^y \\ 0 & \pm J_{yz}^y  & 0 \\ \end{array} \right] \ , \end{eqnarray}
\begin{eqnarray}
{\bf J}_\pm^z &=& \left[ \begin{array} {c c c} \pm J_{xx}^z & 0 & -D_y^z \\
0 & \pm J_{yy}^z & 0 \\ D_y^z & 0 & \pm J_{zz}^z \\ \end{array} \right] \ .
\end{eqnarray}

\begin{table}
\caption{\label{BASIS}Basis spin functions for sites \#1 and \#2 in the unit cell
in terms of the complex-valued Fourier components $S_\alpha (q)$ for irreps
characterized by the eigenvalues of $m_y$ and the glide operation $m_z$.}

\vspace{0.2 in}
\begin{tabular} {||c | c |c || c |c||} 
\hline $\Gamma$ & $m_y$ & $m_z$ & ${\bf S}$(\#1) & ${\bf S}$(\#2) \\ \hline
$\Gamma_1$ & $+$ & $+$ & $(S_x(q),0,S_z(q))$ & $(S_x(q),0,-S_z(q))$ \\
$\Gamma_2$ & $+$ & $-$ & $(S_x(q),0,S_z(q))$ & $(-S_x(q),0,-S_z(q))$ \\
$\Gamma_3$ & $-$ & $+$ & $(0,S_y(q),0)$ & $(0,S_y(q),0)$ \\
$\Gamma_4$ & $-$ & $-$ & $(0,S_y(q),0)$ & $(0,-S_y(q),0)$ \\ \hline
\end{tabular}
\end{table}

\begin{figure}
\includegraphics[width=8cm]{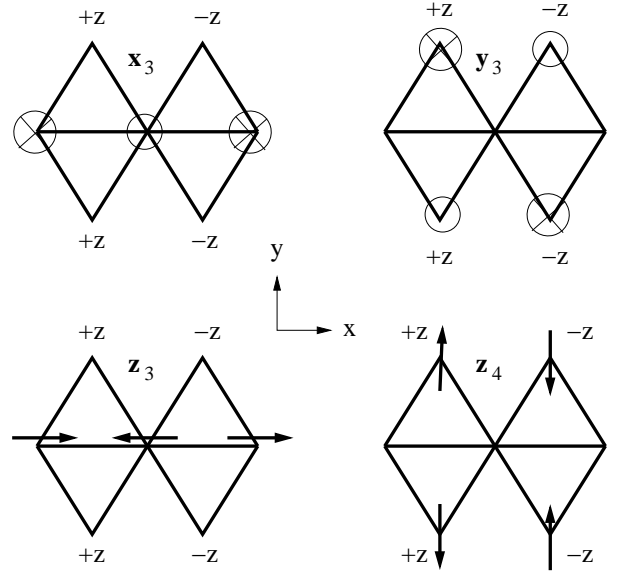}
\caption{\label{TOYPH} Generalized displacements which transform like vectors.
Open circles (circles with inscribed "x") are displacements out of (into) the page.
The sites at positive or negative $z$ are indicated. The upper panels show
modes in which the atoms move only in the $z$-direction.
Upper left: an $x$-like mode $x_3$.  Upper right: a $y$-like mode $y_3$.
The lower panels show $z$-like modes.  Left: $z_3$ with motion only along the
$x$-axis and right: $z_4$ with motion only along the $y$-axis.}
\end{figure}

Next we characterize the spin structure.  There are four irreps for which
the basis functions are listed in Table \ref{BASIS}.  Thus we write
for spins \#1 (${\bf S}_1)$ and \#2 $({\bf S}_2)$ in the unit cell,
\begin{eqnarray}
S_{1,x}(X) &=& [S_x^{(1)}(q) + S_x^{(2)}(q)] e^{iqX} \nonumber \\
&& \ +
[S_x^{(1)}(q)^* + S_x^{(2)}(q)^*] e^{-iqX}  \ , \\
S_{1,y}(X) &=& [S_y^{(3)}(q) + S_y^{(4)}(q)] e^{iqX} \nonumber \\
&& \ + [S_y^{(3)}(q)^* + S_y^{(4)}(q)^*] e^{-iqX}  \ , \\
S_{1,z}(X) &=& [S_z^{(1)}(q) + S_z^{(2)}(q)] e^{iqX} \nonumber \\
&& \ + [S_z^{(1)}(q)^* + S_z^{(2)}(q)^*] e^{-iqX}  \ , \\
S_{2,x}(X) &=& [S_x^{(1)}(q) - S_x^{(2)}(q)] e^{iqX} \nonumber \\
&& \ + [S_x^{(1)}(q)^* - S_x^{(2)}(q)^*] e^{-iqX}  \ , \\
S_{2,y}(X) &=& [S_y^{(3)}(q) - S_y^{(4)}(q)] e^{iqX} \nonumber \\
&& \ + [S_y^{(3)}(q)^* - S_y^{(4)}(q)^*] e^{-iqX}  \ , \\
S_{2,z}(X) &=& [-S_z^{(1)}(q) + S_z^{(2)}(q)] e^{iqX} \nonumber \\
&& \ + [-S_z^{(1)}(q)^* + S_z^{(2)}(q)^*] e^{-iqX}  \ ,
\label{SPINOPS} \end{eqnarray}
where the superscript labels the irrep as in Table \ref{BASIS}).
Thereby we find the trilinear spin-phonon coupling 
(when the spin operators are replaced by their values:
\begin{eqnarray}
V &=& 4N \sin(qa) \sum_p \biggl[ U Q_{x_p} J_{xz}^{x_p}
+ Q_{y_p} (V J_{yz}^{y_p} + W D_z^{y_p} ) \nonumber \\
&& \ + Q_{z_p} \biggl( X D_y^{z_p} + \sum_{\alpha =x,y,z}
Y_\alpha J_{\alpha \alpha}^{z_p} \biggr) \biggr] \ ,
\end{eqnarray}
where
\begin{eqnarray}
U &=& \Im \left( S_x^{(1)}(q)^* S_z^{(1)}(q) + S_x^{(2)}(q) S_z^{(2)}(q)^* \right)
\end{eqnarray}
\begin{eqnarray}
V &=& \Im \left( S_y^{(4)}(q) S_z^{(2)}(q)^* + S_z^{(1)}(q) S_y^{(3)}(q)^* \right)
\end{eqnarray}
\begin{eqnarray}
W &=& \Im \left( S_x^{(1)}(q) S_y^{(3)}(q)^* + S_x^{(2)}(q)^* S_y^{(4)}(q) \right)
\end{eqnarray}
\begin{eqnarray}
X &=& \Im \left( S_x^{(1)}(q)^* S_z^{(2)}(q) + S_x^{(2)}(q) S_z^{(1)}(q)^* \right)
\end{eqnarray}
\begin{eqnarray}
Y_x &=& \Im \left( S_x^{(1)}(q)^* S_x^{(2)}(q) \right)
\end{eqnarray}
\begin{eqnarray}
Y_y &=& \Im \left( S_y^{(3)}(q)^* S_y^{(4)}(q) \right)
\end{eqnarray}
\begin{eqnarray}
Y_z &=& \Im \left( S_z^{(1)}(q) S_z^{(2)}(q)^* \right)
\end{eqnarray}
The general symmetry arguments indicate that there can not be a polarization
along $\hat i$.  We see that $U$ vanishes because all the components within
a single representation have the same phase, so that, for instance,
$S_x^{(1)}(q)S_z^{(1)}(q)^*$ is real.  Here we see that, depending on the
spin structure the spontaneous polarization can either be along
$\hat j$ (if either both irreps \#1 and \#3 are active or both irreps
\#2 and \#4 are active) or along $\hat k$ (if either both irreps \#1 and \#2
are active or both irreps \#3 and \#4 are active.)  These results are
exactly what the phenomenological analysis would give.

\section{Conclusions}
In this paper we present neutron scattering measurements of phonons in NVO,
the first-principles computation of the zone-center phonons and their symmetry
analysis. We identified two particularly interesting phonons among the twelve
B$_{2u}$ modes which have the right symmetry to induce the experimentally
observed dipole moment along the ${\bf b}$-axis in NVO. Using the calculated atomic
charges and the eigenvectors we conclude that the required distortion to
induce the observed dipole moment is small (~0.001 \AA) 
and would be difficult to observe directly by neutron powder diffraction.
Finally, we present 
a symmetry analysis to characterize the microscopic
magnetoelectric coupling in Ni$_3$V$_2$O$_8$.  The method can easily
be applied to similar systems such as TbMnO$_3$.  The result of
this analysis is that we can specify those strain derivatives of
the exchange tensor which should now be the targets of more
fundamental quantum calculations, perhaps based on the LDA\cite{LDA}
or similar schemes.  (These results are given in Eqs. (\ref{G1EQ})-(\ref{G3EQ}),
where the superscript indicates the component of the gradient.)  
In Subsec. \ref{RESULTS} we show how these gradients lead to
an evaluation of the spontaneous polarization from first principles. 
In Sec. \ref{TOYSEC} we also studied some structurally simpler
toy models.  A general conclusion is that the local site symmetry
in NVO is low enough that almost all strain derivatives of the
exchange tensor are involved.  It is true that the Dzyaloshinskii-Moriya
interaction must be present (and, of course, it is allowed in such
low symmetry systems).  In the presence of such interactions ferroelectricity
results from the strain dependence of all the components of the
exchange tensor, even that of the isotropic exchange interaction.

\begin{acknowledgments}
This work was supported in part by the Israel US Binational Science
Foundation under grant number 2000073.
 We thank C. Broholm for providing us
the NVO powder samples used in this study.
\end{acknowledgments}

\begin{appendix}
\section{Next-nearest neighbor spine interactions}

In this section we consider next-nearest neighbor (nnn) interactions
between spins on a given spine.  Since only the gradients with respect 
to $Q_{y_p}$ are needed, we only consider those here.  We set
\begin{eqnarray}
{\partial J_{\alpha \beta} (1,1') \over \partial Q_{y_p}}
&=& V_{\alpha \beta}^{y_p} \ .
\end{eqnarray}
The operation $2_y$ takes this bond into itself with reversed indices.
So, taking account of the transformation properties of $Q_{y_p}$,
we require that
\begin{eqnarray} {\bf 2}_y {\bf V}^{y_p} &=& \tilde {\bf V}^{y_p} \ ,
\end{eqnarray}
where ${\bf 2}_y$ is the two-fold rotation operator. In terms of matrices,
this relation is
\begin{eqnarray}
2_y {\bf V}^{y_p} 2_y &=& \tilde {\bf V}^{y_p} \ ,
\end{eqnarray}
where $2_y = \sigmav_x \sigmav_z$.  Thus
\begin{eqnarray}
{\bf V}^{y_p} &=& \left[ \begin{array} {c c c}
K_{xx}^y & E_z^y & K_{xz}^y \\ - E_z^y & K_{yy}^y & E_x^y \\
K_{xz}^y & - E_x^y & K_{zz}^y \\ \end{array} \right] \ .
\end{eqnarray}
We have that
\begin{eqnarray}
{\partial J_{\alpha \beta} (2,2') \over \partial Q_{y_p}} &=&
- 2_x {\bf V}^{y_p} 2_x \ ,
\end{eqnarray}
so that
\begin{eqnarray}
{\partial J_{\alpha \beta} (2,2') \over \partial Q_{y_p}} &=&
\left[ \begin{array} {c c c}
-K_{xx}^y & E_z^y & K_{xz}^y \\ - E_z^y & -K_{yy}^y & -E_x^y \\
K_{xz} & E_x^y & -K_{zz}^y \\
\end{array} \right] \ .
\end{eqnarray}
We have that
\begin{eqnarray}
{\partial J_{\alpha \beta} (3',3) \over \partial Q_{y_p}} &=&
- \sigmav_y {\bf V}^{\gamma_p} \sigmav_y \ ,
\end{eqnarray}
so that
\begin{eqnarray}
{\partial J_{\alpha \beta} (3',3) \over \partial Q_{y_p}} &=&
\left[ \begin{array} {c c c}
-K_{xx}^y & E_z^y & - K_{xz}^y \\ -E_z^y & -K_{yy}^y & E_x^y \\ -K_{xz}^y & -E_x^y & -K_{zz}^y \\
\end{array} \right] \ .
\end{eqnarray}
We have that
\begin{eqnarray}
{\partial J_{\alpha \beta} (4',4) \over \partial Q_{y_p}} &=&
\sigmav_x \tilde {\bf V}^{\gamma_p} \sigmav_x \ ,
\end{eqnarray}
so that
\begin{eqnarray}
{\partial J_{\alpha \beta} (4',4) \over \partial Q_{y_p}} &=&
\left[ \begin{array} {c c c}
K_{xx}^y & E_z^y & - K_{xz}^y \\ - E_z^y & K_{yy}^y & - E_x^y \\
-K_{xz}^y & E_x^y & K_{zz}^y \\
\end{array} \right] \ .
\end{eqnarray}

\section{Spine cross-tie interactions}

\begin{figure}
\includegraphics[height=5.5cm]{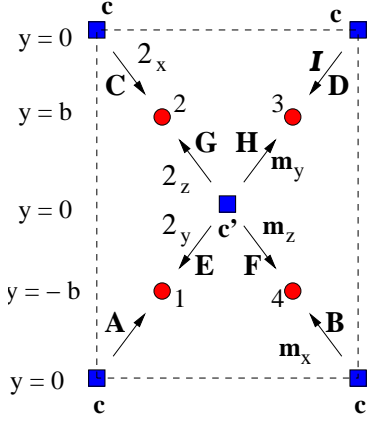}
\caption{\label{SPCR} As Fig. \protect{\ref{SPSP}}.  Here we show
the eight different nearest-neighbor spine-cross tie interactions,
which we label A ... H.  The arrow points from the first site index
to the second site index. We also give the symmetry operation one
has to apply to interaction A to get each of the other interactions.}
\end{figure}

In this section we analyze the spine-cross tie interactions, 
in Fig. \ref{SPCR}, again keeping only derivative with respect to
$Q_{y_p}$.  We set the interaction of type A to be
\begin{eqnarray}
{\partial J_{\alpha \beta} (c,1) \over dQ_{y_p}} &=&
W_{\alpha \beta}^{y_p} \ ,
\end{eqnarray}
so that (when the index $p$ is suppressed)
\begin{eqnarray}
{\bf W}^y &=& \left[ \begin{array} {c c c}
L_{xx}^y & L_{xy}^y + F_z^y & L_{xz}^y - F_y^y \\
L_{xy}^y - F_z^y & L_{yy}^y & L_{yz}^y + F_x^y \\
L_{xz}^y + F_y^y & L_{yz}^y - F_x^y & L_{zz}^y \\
\end{array} \right] \ .
\end{eqnarray}

\noindent
We have that
\begin{eqnarray}
{\partial {\bf J} (c,4) \over \partial Q_{y_p}} &=&
\sigmav_x  {\bf W}^{\gamma_p} \sigmav_x \ ,
\end{eqnarray}
so that
\begin{eqnarray}
{\partial {\bf J} (c,4) \over \partial Q_{y_p}} &=&
\left[ \begin{array} {c c c}
L_{xx}^y & - L_{xy}^y - F_z^y & - L_{xz}^y + F_y^y \\
- L_{xy}^y + F_z^y & L_{yy}^y & L_{yz}^y + F_x^y \\
- L_{xz}^y - F_y^y & L_{yz}^y - F_x^y & L_{zz}^y \\
\end{array} \right] \ . \nonumber \\ 
\end{eqnarray}
We have that
\begin{eqnarray}
{\partial {\bf J} (c,2) \over \partial Q_{y_p}} &=&
- 2_x  {\bf W}^{y_p} 2_x \ ,
\end{eqnarray}
so that
\begin{eqnarray}
{\partial {\bf J} (c,2) \over \partial Q_{y_p}} &=&
\left[ \begin{array} {c c c}
- L_{xx}^y & L_{xy}^y + F_z^y & L_{xz}^y - F_y^u \\
L_{xy}^y - F_z^y & - L_{yy}^y & - L_{yz}^y - F_x^y \\
L_{xz}^y + F_y^y & -L_{yz}^y + F_x^y & - L_{zz}^y \\
\end{array} \right] \ . \nonumber \\
\end{eqnarray}
We have that
\begin{eqnarray}
{\partial {\bf J} (c,3) \over \partial Q_{y_p}} &=&
- {\cal I}  {\bf W}^{\gamma_p} {\cal I} \ ,
\end{eqnarray}
so that
\begin{eqnarray}
{\partial {\bf J} (c,3) \over \partial Q_{y_p}} &=&
\left[ \begin{array} {c c c}
- L_{xx}^y & -L_{xy}^y - F_z^y & -L_{xz}^y + F_y^y \\
- L_{xy}^y + F_z^y & - L_{yy}^y & - L_{yz}^y - F_x^y \\
- L_{xz}^y - F_y^y & - L_{yz}^y + F_x^y & - L_{zz}^y \\
\end{array} \right] \ . \nonumber \\
\end{eqnarray}
We have that
\begin{eqnarray}
{\partial {\bf J} (c',1) \over \partial Q_{y_p}} &=&
2_y {\bf W} 2_y \ ,
\end{eqnarray}
so that
\begin{eqnarray}
{\partial {\bf J} (c',1) \over \partial Q_{y_p}} &=&
\left[ \begin{array} {c c c}
L_{xx}^y & -L_{xy}^y - F_z^y &  L_{xz}^y - F_y^y \\
-L_{xy}^y + F_z^y &  L_{yy}^y & -L_{yz}^y - F_x^y \\
 L_{xz}^y + F_y^y & -L_{yz}^y + F_x^y &  L_{zz}^y \\
\end{array} \right] \ . \nonumber \\
\end{eqnarray}
We have that
\begin{eqnarray}
{\partial {\bf J} (c',4) \over \partial Q_{y_p}} &=&
\sigmav_z  {\bf W}^{\gamma_p} \sigmav_z \ ,
\end{eqnarray}
so that
\begin{eqnarray}
{\partial {\bf J} (c',4) \over \partial Q_{y_p}} &=&
\left[ \begin{array} {c c c}
L_{xx}^y & L_{xy}^y + F_z^y & - L_{xz}^y+ F_y^y \\
L_{xy}^y - F_z^y & L_{yy}^y & - L_{yz}^y - F_x^y \\
-L_{xz}^y - F_y^y & - L_{yz}^y + F_x^y & L_{zz}^y \\
\end{array} \right] \ . \nonumber \\
\end{eqnarray}
We have that
\begin{eqnarray}
{\partial {\bf J} (c',2) \over \partial Q_{y_p}} &=&
- 2_z  {\bf W}^{\gamma_p} 2_z \ ,
\end{eqnarray}
so that
\begin{eqnarray}
{\partial {\bf J} (c',2) \over \partial Q_{y_p}} &=&
\left[ \begin{array} {c c c}
- L_{xx}^y & - L_{xy}^y - F_z^y & L_{xz}^y - F_y^y \\
- L_{xy}^y + F_z^y & - L_{yy}^y & L_{yz}^y + F_x^y \\
L_{xz}^y + F_y^y & L_{yz}^y - F_x^y & - L_{zz}^y \\
\end{array} \right] \ . \nonumber \\
\end{eqnarray}
We have that
\begin{eqnarray}
{\partial {\bf J} (c',3) \over \partial Q_{y_p}} &=&
- \sigmav_y  {\bf W}^{\gamma_p} \sigmav_y \ ,
\end{eqnarray}
so that
\begin{eqnarray}
{\partial {\bf J} (c',3) \over \partial Q_{y_p}} &=&
\left[ \begin{array} {c c c}
- L_{xx}^y & L_{xy}^y + F_z^y & - L_{xz}^y + F_y^y \\
L_{xy}^y - F_z^y & - L_{yy}^y & L_{yz}^y + F_x^y \\
- L_{xz}^y - F_y^y & L_{yz}^y - F_x^y & - L_{zz}^y \\
\end{array} \right] \ . \nonumber \\
\end{eqnarray}

\section{Spin-Phonon Interaction}

\subsection{nn spine interactions}

Now we collect terms involving $J_{xx}^y$:
\begin{eqnarray}
V_y(J_{xx}^y) &=& Q_yJ_{xx}^y \sum_{uc} \biggl[
S_x(1) S_x(4) - S_x(2) S_x(3) \nonumber \\ && \
+ S_x(4) S_x(1') - S_x(3) S_x(2') \biggr]
\nonumber \\ &=& Q_y J_{xx}^y \sum_{uc} e^{iqx_{14}} \biggl[
(a_x+ib_x)(-a_x^*+ib_x^*) \nonumber \\ && \
- (-a_x+ib_x)(a_x^*+ib_x^*) \nonumber \\ && \
+ (-a_x-ib_x)(a_x^*-ib_x^*) \nonumber \\ && \
- (a_x-ib_x)(-a_x^*-ib_x^*) \biggr] + {\rm c. \ c.}
\nonumber \\ &=& 4Q_y J_{xx}^y N_{uc} \left( e^{iqx_{14}} [
ia_xb_x^* -i b_xa_x^*]  + {\rm c. \ c.} \right) 
\nonumber \\ &=&
- 16 N_{uc} Q_y J_{xx}^y \cos(qa/2) \Im [ a_x b_x^* ] \ .
\end{eqnarray}

Now we collect terms involving $J_{yy}^y$:
\begin{eqnarray}
V_y(J_{yy}^y) &=& Q_yJ_{yy}^y \sum_{uc} \biggl[
S_y(1) S_y(4) - S_y(2) S_y(3) \nonumber \\ && \
+ S_y(4) S_y(1') - S_y(3) S_y(2') \biggr]
\nonumber \\ &=& Q_y J_{yy}^y \sum_{uc} e^{iqx_{14}} \biggl[
(ia_y+b_y)(ia_y^*-b_y^*) \nonumber \\ && \
- (ia_y-b_y)(ia_y^*+b_y^*) \nonumber \\ && \
+ (-ia_y-b_y)(-ia_y^*+b_y^*) \nonumber \\ && \
- (-ia_y+b_y)(-ia_y^*-b_y^*) \biggr]
+ {\rm c. \ c.} \nonumber \\ &=& 4Q_y J_{yy}^y N_{uc} \left( e^{iqx_{14}} [
-ia_yb_y^* + i b_ya_y^*]  + {\rm c. \ c.} \right) \nonumber
\\ &=& 16 N_{uc} Q_y J_{yy}^y \cos(qa/2) \Im [a_yb_y^*] \ .
\end{eqnarray}

Now we collect terms involving $J_{zz}^y$:
\begin{eqnarray}
V_y (J_{zz}^y) &=& Q_yJ_{zz}^y \sum_{uc} \biggl[
S_z(1) S_z(4) - S_z(2) S_z(3) \nonumber \\ && \
+ S_z(4) S_z(1') - S_z(3) S_z(2') \biggr]
\nonumber \\ &=& Q_y J_{zz}^y \sum_{uc} e^{iqx_{14}} \biggl[
(a_z+ib_z)(a_z^*-ib_z^*) \nonumber \\ && \
- (a_z-ib_z)(a_z^*+ib_z^*) \nonumber \\ && \
+ (a_z+ib_z)(a_z^*-ib_z^*)  \nonumber \\ && \
- (a_z-ib_z)(a_z^*+ib_z^*) \biggr] + {\rm c. \ c.}
\nonumber \\ &=& 4Q_y J_{zz}^y N_{uc} \left( e^{iqx_{14}} [
-ia_zb_z^* + i b_za_z^*]  + {\rm c. \ c.} \right) \nonumber
\\ &=& 16 N_{uc} Q_y J_{zz}^y \cos (qa/2) \Im [a_z b_z^* ] .
\end{eqnarray}

Now we collect terms involving $J_{yz}^y$:
\begin{eqnarray}
V_y(J_{yz}^y) &=& Q_yJ_{yz}^y \sum_{uc} \biggl[
S_y(1) S_z(4) + S_z(1) S_y(4) \nonumber \\ && \
- S_z(2) S_y(3) - S_y(2) S_z(3) \nonumber \\
&& \ - S_y(4) S_z(1') - S_z(4) S_y(1') \nonumber \\ && \
+  S_y(3) S_z(2') + S_z(3) S_y(2')
\biggr] \nonumber \\ &=& Q_y J_{yz}^y \sum_{uc} e^{iqx_{14}} \biggl[
(ia_y+b_y)(a_z^*-ib_z^*) \nonumber \\ && \
+ (a_z+ib_z)(ia_y^*-b_y^*) \nonumber \\ &&
- (a_z-ib_z)(ia_y^*+b_y^*)  \nonumber \\ && \
- (ia_y-b_y)(a_z^*+ib_z^*) \nonumber \\ && \
- (-ia_y -b_y)(a_z^*-ib_z^*) \nonumber \\ && \
 - (a_z+ib_z)(-ia_y^*+b_y^*) \nonumber \\ && \
+ (-ia_y+b_y)(a_z^*+ib_z^*) \nonumber \\ && \
 + (a_z-ib_z)(-ia_y^*-b_y^*) \biggr] + {\rm c. \ c.}
\nonumber \\ &=& 4Q_y J_{yz}^y N_{uc} \left( e^{iqx_{14}} [
a_yb_z^* + b_ya_z^* - a_zb_y^* - b_za_y^* ] \right.
\nonumber \\ && \ \left.
+ {\rm c. \ c.} \right) \nonumber \\
&=& -16 N_{uc} Q_y \sin(qa/2) J_{yz}^y 
\Im [a_y^*b_z+b_y^*a_z] \ .
\end{eqnarray}

Now we collect terms proportional to $D_z^y$:
\begin{eqnarray}
V_y(D_z^y) &=& Q_y D_z^y \sum_{uc} \biggl[
S_x(1)S_y(4)-S_y(1)S_x(4) \nonumber \\ && \
+ S_x(2)S_y(3) - S_y(2)S_x(3) \nonumber \\ &&
+ S_x(4)S_y(1') - S_y(4)S_x(1') \nonumber \\ && \
+ S_x(3)S_y(2') - S_y(3)S_x(2') \biggr]
\nonumber \\ &=& Q_yD_z^y \sum_{uc} e^{iqx_{14}} \biggl[
(a_x+ib_x)(ia_y^*-b_y^*) \nonumber \\ && \
 - (ia_y+b_y)(-a_x^*+ib_x^*) \nonumber \\ && \
+ (-a_x+ib_x)(ia_y^*+b_y^*) \nonumber \\ && \
 - (ia_y-b_y)(a_x^*+ib_x^*) \nonumber \\ && \
+ (-a_x-ib_x)(-ia_y^*+b_y^*) \nonumber \\ && \
 - (-ia_y-b_y)(a_x^*-ib_x^*) \nonumber \\ && \
+ (a_x-ib_x)(-ia_y^*-b_y^*) \nonumber \\ && \
 - (-ia_y+b_y)(-a_x^*-ib_x^*) \biggr] + 
{\rm c. \ c.} \nonumber \\ &=& \
4Q_y D_z^y N_{uc} \left( e^{iqx_{14}} [-a_xb_y^*-b_xa_y^*+a_yb_x^*+b_ya_x^*]
\right. \nonumber \\ && \left. \ + {\rm c. \ c.} \right) \nonumber \\ &=&
-16 N_{uc} Q_y \sin(qa/2) D_z^y \Im [a_y^*b_x+b_y^*a_x] \ .
\end{eqnarray}

Now we collect terms proportional to $D_y^y$:
\begin{eqnarray}
V_y(D_y^y)  &=& Q_y D_y^y \sum_{uc} \biggl[
- S_x(1)S_z(4) + S_z(1)S_x(4) \nonumber \\ && \
 - S_x(2)S_z(3) + S_z(2)S_x(3) \nonumber \\ &&
+ S_x(4)S_z(1') - S_z(4)S_x(1') \nonumber \\ && \
 + S_x(3)S_z(2') - S_z(3)S_x(2') \biggr]
\nonumber \\ &=& Q_yD_y^y \sum_{uc} e^{iqx_{14}} \biggl[
-(a_x+ib_x)(a_z^*-ib_z^*) \nonumber \\ && \
 + (a_z+ib_z)(-a_x^*+ib_x^*) \nonumber \\ && \
- (-a_x+ib_x)(a_z^*+ib_z^*) \nonumber \\ && \
 + (a_z-ib_z)(a_x^*+ib_x^*) \nonumber \\ && \
+ (-a_x-ib_x)(a_z^*-ib_z^*) \nonumber \\ && \
 - (a_z+ib_z)(a_x^*-ib_x^*) \nonumber \\ && \
+ (a_x-ib_x)(a_z^*+ib_z^*) \nonumber \\ && \
 - (a_z-ib_z)(-a_x^*-ib_x^*) \biggr] \nonumber
\\ && \ + {\rm c. \ c.} \nonumber \\ &=& \
4Q_y D_y^y N_{uc} \left( e^{iqx_{14}} i[a_xb_z^* - b_x a_z^* + a_zb_x^* -b_za_x^*]
\right. \nonumber \\&& \left. \ + {\rm c. \ c.} \right) \nonumber \\ &=&
-16 N_{uc} Q_y \cos(qa/2) D_y^y \Im [a_xb_z^*+b_x^*a_z] \ .
\end{eqnarray}

\subsection{nnn spine interactions}

First we consider terms proportional to $K_{xx}^y$ (where we omit
the $p$ index):
\begin{eqnarray}
V_y(K_{xx}^y) &=& Q_y K_{xx}^y \sum_{uc} [ S_x(1) S_x(1') - S_x(2) S_x(2')
\nonumber \\ && \ - S_x(3) S_x(3') + S_x(4) S_x(4')] \nonumber \\
&=& N_{uc}Q_y K_{xx}^y [ t_{xx} + t_{xx}^*] \ ,
\end{eqnarray}
where
\begin{eqnarray}
t_{xx} &=& e^{-iqa} \bigl[
(a_{s,x} + i b_{s,x}) (a_{s,x}^*-ib_{s,x}^*)
\nonumber \\ && \
- (-a_{s,x} + i b_{s,x}) (-a_{s,x}^*-ib_{s,x}^*) \nonumber \\ && \ 
-(a_{s,x} - i b_{s,x}) (a_{s,x}^* + ib_{s,x}^*)
\nonumber \\ && \
+ (-a_{s,x} - i b_{s,x}) (-a_{s,x}^* +ib_{s,x}^*) \bigr]
\nonumber \\ &=&
e^{-iqa} \bigl[ -4i a_{s,x} b_{s,x}^* + 4i a_{s,x}^* b_{s,x} \bigr]
\nonumber \\ &=& 8 e^{-iqa} \Im [a_{s,x} b_{s,x}^* ] \ ,
\end{eqnarray}
so that
\begin{eqnarray}
V_y(K_{xx}^y) &=& 16 N_{uc} Q_y K_{xx}^y \cos (qa) \Im [a_{s,x} b_{s,x}^*] \ . 
\end{eqnarray}

Next, we consider terms proportional to $K_{yy}^y$:
\begin{eqnarray}
V_y(K_{yy}^y) &=& Q_y K_{yy}^y \sum_{uc} [ S_y(1) S_y(1') - S_y(2) S_y(2')
\nonumber \\ && \ - S_y(3) S_y(3') + S_y(4) S_y(4') ] \nonumber \\
&=& N_{uc}Q_y K_{yy}^y [ t_{yy} + t_{yy}^*] \ ,
\end{eqnarray}
where
\begin{eqnarray}
t_{yy} &=& e^{-iqa} \bigl[
(ia_{s,y} + b_{s,y}) (-ia_{s,y}^* +b_{s,y}^*)
\nonumber \\ && \
- (ia_{s,y} - b_{s,y}) (-ia_{s,y}^*-b_{s,y}^*) \nonumber \\ && \ 
-(-ia_{s,y} + b_{s,y}) (ia_{s,y}^* + b_{s,y}^*)
\nonumber \\ && \
+ (-ia_{s,y} - b_{s,y}) (ia_{s,y}^* -b_{s,y}^*) \bigr]
\nonumber \\ &=&
e^{-iqa} \bigl[ 4i a_{s,y} b_{s,y}^* - 4i a_{s,y}^* b_{s,y} \bigr]
\nonumber \\ &=& 8 e^{-iqa} \Im [a_{s,y}^* b_{s,y} ] \ ,
\end{eqnarray}
so that
\begin{eqnarray}
V_y(K_{yy}^y) &=& 16 N_{uc} Q_y K_{yy}^y \cos (qa) \Im [a_{s,y}^* b_{s,y}] \ . 
\end{eqnarray}
 
Next,  we consider terms proportional to $K_{zz}^y$:
\begin{eqnarray}
V_y(K_{zz}^y) &=& Q_y K_{zz}^y \sum_{uc} \bigl( S_z(1) S_z(1') - S_z(2) S_z(2')
\nonumber \\ && \ - S_z(3) S_z(3') + S_z(4) S_z(4') \bigr) \nonumber \\
&=& N_{uc}Q_y K_{zz}^y [ t_{zz} + t_{zz}^*] \ ,
\end{eqnarray}
where
\begin{eqnarray}
t_{zz} &=& e^{-iqa} \bigl[
(a_{s,z} + ib_{s,z}) (a_{s,z}^* - ib_{s,z}^*)
\nonumber \\ && \
- (a_{s,z} - ib_{s,z}) (a_{s,z}^* +ib_{s,z}^*) \nonumber \\ && \ 
-(a_{s,z} -i b_{s,z}) (a_{s,z}^* + ib_{s,z}^*) \nonumber \\ && \
+ (a_{s,z} +i b_{s,z}) (a_{s,z}^* - ib_{s,z}^*) \bigr]
\nonumber \\ &=&
e^{-iqa} \bigl[ -4i a_{s,z} b_{s,z}^* + 4i a_{s,z}^* b_{s,z} \bigr]
\nonumber \\ &=& 8 e^{-iqa} \Im [a_{s,z} b_{s,z}^* ] \ ,
\end{eqnarray}
so that
\begin{eqnarray}
V_y(K_{zz}^y) &=& 16 N_{uc} Q_y K_{zz}^y \cos (qa) \Im [a_{s,z} b_{s,z}^*] \ . 
\end{eqnarray}

Now we consider terms proportional to $E_z^y$:
\begin{eqnarray}
V_y(E_z^y) &=& Q_y E_z^y \sum_{uc} \bigl[ S_x(1) S_y(1') - S_y(1) S_x(1')
\nonumber \\ && \ + S_x(2) S_y(2') - S_y(2) S_x(2') \nonumber \\ && \ 
+ S_x(3') S_y(3) - S_y(3') S_x(3) \nonumber \\ && \
+ S_x(4') S_y (4) - S_y(4') S_x(4) \bigr] \nonumber \\
&=& N_{uc}Q_y E_z^y [ t_{zy} + t_{zy}^*] \ ,
\end{eqnarray}
where
\begin{eqnarray}
t_{zy} &=& e^{-iqa} \bigl[
(a_{s,x} + ib_{s,x}) (-ia_{s,y}^*  + b_{s,y}^*) \nonumber \\ && \
- (ia_{s,y} + b_{s,y}) (a_{s,x}^* - ib_{s,x}^*) \nonumber \\ && \
+ (-a_{s,x} +i b_{s,x}) (-ia_{s,y}^* - b_{s,y}^*) \nonumber \\ && \
- (ia_{s,y} - b_{s,y}) (-a_{s,x}^* - ib_{s,x}^*) \nonumber \\ && \
+ (a_{s,x} - ib_{s,x}) (ia_{s,y}^*  + b_{s,y}^*) \nonumber \\ && \
- (-ia_{s,y} + b_{s,y}) (a_{s,x}^* + ib_{s,x}^*) \nonumber \\ && \
+ (-a_{s,x} - i b_{s,x}) (ia_{s,y}^* - b_{s,y}^*) \nonumber \\ && \
- (-ia_{s,y} - b_{s,y}) (-a_{s,x}^* + ib_{s,x}^*) \bigr] \nonumber \\ &=&
4e^{-iqa} \bigl[ a_{s,x} b_{s,y}^* +  b_{s,x} a_{s,y}^*
- a_{s,y} b_{s,x}^* - b_{s,y}a_{s,x}^*  \bigr]
\nonumber \\ &=& 8 i e^{-iqa} \Im [a_{s,x} b_{s,y}^*+b_{s,x} a_{s,y}^* ] \ ,
\end{eqnarray}
so that
\begin{eqnarray}
V_y(E_z^y) &=& 16 N_{uc} Q_y E_z^y \sin (qa) \Im [a_{s,x} b_{s,y}^*
+ b_{s,x}a_{s,y}^* ] \ . 
\end{eqnarray}

Now we consider terms proportional to $E_x^y$:
\begin{eqnarray}
V_y(E_x^y) &=& Q_y E_x^y \sum_{uc} \bigl[ S_y(1) S_z(1') - S_z(1) S_y(1')
\nonumber \\ && \
- S_y(2) S_z(2') + S_z(2) S_y(2') \nonumber \\ && \ 
+ S_y(3') S_z(3) - S_z(3') S_y(3) \nonumber \\ && \
- S_y(4') S_z (4) + S_z(4') S_y(4) \bigr] \nonumber \\
&=& N_{uc}Q_y E_x^y [ t_{xy} + t_{xy}^*] \ ,
\end{eqnarray}
where
\begin{eqnarray}
t_{xy} &=& e^{-iqa} \bigl[
(ia_{s,y} + b_{s,y}) (a_{s,z}^*  -i b_{s,z}^*) \nonumber \\ && \
- (a_{s,z} + ib_{s,z}) (-ia_{s,y}^*  + b_{s,y}^*) \nonumber \\ && \
- (ia_{s,y} - b_{s,y}) (a_{s,z}^* +i b_{s,z}^*) \nonumber \\ && \
+ (a_{s,z} - ib_{s,z}) (-ia_{s,y}^* - b_{s,y}^*) \nonumber \\ &&
+ (-ia_{s,y} + b_{s,y}) (a_{s,z}^*  + ib_{s,z}^*) \nonumber \\ && \
- (a_{s,z} -i b_{s,z}) (ia_{s,y}^* + b_{s,y}^*) \nonumber \\ && \
- (-ia_{s,y} - b_{s,y}) (a_{s,z}^* -i b_{s,z}^*) \nonumber \\ && \
+ (a_{s,z} +i b_{s,z}) (ia_{s,y}^* - b_{s,y}^*) \bigr] \nonumber \\ &=&
4e^{-iqa} \bigl[ a_{s,y} b_{s,z}^* +  b_{s,y} a_{s,z}^* \nonumber \\ && \
- a_{s,z} b_{s,y}^* - b_{s,z}a_{s,y}^*  \bigr]
\nonumber \\ &=& 8 i e^{-iqa} \Im [a_{s,y} b_{s,z}^*+b_{s,y} a_{s,z}^* ] \ ,
\end{eqnarray}
so that
\begin{eqnarray}
V_y(E_x^y) &=& 16 N_{uc} Q_y E_x^y \sin (qa) \Im [a_{s,y} b_{s,z}^*
+ b_{s,y}a_{s,z}^* ] \ . 
\end{eqnarray}

Now we consider terms proportional to $K_{xz}^y$:
\begin{eqnarray}
V_y(K_{xz}^y) &=& Q_y K_{xz}^y \sum_{uc} \bigl[ S_x(1) S_z(1') + S_z(1) S_x(1')
\nonumber \\ && \
+ S_x(2) S_z(2') + S_z(2) S_x(2') \nonumber \\ && \ 
- S_x(3) S_z(3') - S_z(3) S_x(3')\nonumber \\ && \
- S_x(4) S_z (4') - S_z(4) S_x(4')
\bigr] \nonumber \\
&=& N_{uc}Q_y K_{xz}^y [ t_{xz} + t_{xz}^*] \ ,
\end{eqnarray}
where
\begin{eqnarray}
t_{xz} &=& e^{-iqa} \bigl[
(a_{s,x} + ib_{s,x}) (a_{s,z}^*  -i b_{s,z}^*) \nonumber \\ && \
+ (a_{s,z} + ib_{s,z}) (a_{s,x}^*  -i b_{s,x}^*) \nonumber \\ && \
+ (-a_{s,x} +i b_{s,x}) (a_{s,z}^* +i b_{s,z}^*) \nonumber \\ && \
+ (a_{s,z} - ib_{s,z}) (-a_{s,x}^* - ib_{s,x}^*) \nonumber \\ &&
- (a_{s,x} -i b_{s,x}) (a_{s,z}^*  + ib_{s,z}^*) \nonumber \\ && \
- (a_{s,z} -i b_{s,z}) (a_{s,x}^* + ib_{s,x}^*) \nonumber \\ && \
- (-a_{s,x} - ib_{s,x}) (a_{s,z}^* -i b_{s,z}^*) \nonumber \\ && \
- (a_{s,z} +i b_{s,z}) (-a_{s,y}^* +i b_{s,y}^*) \bigr] \nonumber \\ &=&
4ie^{-iqa} \bigl[ -a_{s,x} b_{s,z}^* +  b_{s,x} a_{s,z}^*
- a_{s,z} b_{s,x}^* + b_{s,z}a_{s,x}^*  \bigr]
\nonumber \\ &=& 8 e^{-iqa} \Im [a_{s,x} b_{s,z}^*+a_{s,z} b_{s,x}^* ] \ ,
\end{eqnarray}
so that
\begin{eqnarray}
V_y(K_{xz}^y) &=& 16 N_{uc} Q_y K_{xz}^y \cos (qa) \nonumber \\ && \
\times \Im [a_{s,x} b_{s,z}^* + a_{s,z} b_{s,x}^* ] \ . 
\end{eqnarray}

\subsection{Spine cross-tie interactions}

Here we analyze the spin-cross tie interactions.

We consider the terms proportional to $L_{xx}^y$:
\begin{eqnarray}
V_y(L_{xx}^y) &=& Q_y L_{xx}^y [ S_x(5) + S_x(6)]
\nonumber \\ && \ \times
[S_x(1) - S_x(2) - S_x(3) + S_x(4)] \nonumber \\
&\equiv& N_{uc} Q_y L_{xx}^y [ u_{xx} + u_{xx}^*] \ ,
\end{eqnarray}
where
\begin{eqnarray}
u_{xx} &=& b_{cx} [ (a_{s,x}^*-ib_{s,x}^*)e^{-iqa/4} \nonumber \\ && \
- (-a_{s,x}^*-ib_{s,x}^*)e^{-iqa/4}
- (a_{s,x}^*+ib_{s,x}^*)e^{iqa/4} \nonumber \\ && \
+ (-a_{s,x}^*+ib_{s,x}^*)e^{iqa/4} \nonumber \\ && \
- (a_{s,x}^*-ib_{s,x}^*)e^{iqa/4}
+ (-a_{s,x}^*-ib_{s,x}^*)e^{iqa/4} \nonumber \\ && \
+ (a_{s,x}^*+ib_{s,x}^*)e^{-iqa/4}
- (-a_{s,x}^*+ib_{s,x}^*)e^{-iqa/4} ] \nonumber \\ &=& 
4 b_{cx} a_{s,x}^* [e^{-iqa/4} - e^{iqa/4}]
\nonumber \\ &=& -8i \sin(qa/4) b_{cx} a_{s,x}^*
\end{eqnarray}
so that
\begin{eqnarray}
V_y(L_{xx}^y) &=& 16 N_{uc} Q_y L_{xx}^y \sin (qa/4) \Im [b_{cx} a_{s,x}^*] \ .
\end{eqnarray}

We consider the terms proportional to $L_{yy}^y$:
\begin{eqnarray}
V_y(L_{yy}^y) &=& Q_y L_{yy}^y [ S_y(5) + S_y(6)] \nonumber \\ && \
[S_y(1) - S_y(2) - S_y(3) + S_y(4)] \nonumber \\ &\equiv&
N_{uc} Q_y L_{yy}^y [ u_{yy} + u_{yy}^*] \ ,
\end{eqnarray}
where
\begin{eqnarray}
u_{yy} &=& a_{cy} [ (-ia_{s,y}^*+b_{s,y}^*)e^{-iqa/4}
- (-ia_{s,y}^*-b_{s,y}^*)e^{-iqa/4} \nonumber \\ && \
- (ia_{s,y}^*+b_{s,y}^*)e^{iqa/4}
+ (ia_{s,y}^*-b_{s,y}^*)e^{iqa/4} \nonumber \\ && \
- (-ia_{s,y}^*+b_{s,y}^*)e^{iqa/4}
+ (-ia_{s,y}^*-b_{s,y}^*)e^{iqa/4} \nonumber \\ && \
+ (ia_{s,y}^*+b_{s,y}^*)e^{-iqa/4}
- (ia_{s,y}^*-b_{s,y}^*)e^{-iqa/4} ] \nonumber \\ &=& 
4 a_{cy} b_{s,y}^* [e^{-iqa/4} - e^{iqa/4}]
\nonumber \\ &=& -8i \sin(qa/4) a_{cy} b_{s,y}^*
\end{eqnarray}
so that
\begin{eqnarray}
V_y(L_{yy}^y) &=& 16 N_{uc} Q_y L_{yy}^y \sin (qa/4) \Im [a_{cy} b_{s,y}^*] \ .
\end{eqnarray}

We consider the terms proportional to $L_{zz}^y$:
\begin{eqnarray}
V_y(L_{zz}^y) &=& Q_y L_{zz}^y [ S_z(5) + S_z(6)] \nonumber \\ &&
\ \times [S_z(1) - S_z(2) - S_z(3) + S_z(4)] \nonumber \\
&\equiv& N_{uc} Q_y L_{zz}^y [ u_{zz} + u_{zz}^*] \ ,
\end{eqnarray}
where
\begin{eqnarray}
u_{zz} &=& a_{cz} [ (a_{s,z}^*-ib_{s,z}^*)e^{-iqa/4}
- (a_{s,z}^*+ib_{s,z}^*)e^{-iqa/4} \nonumber \\ && \
- (a_{s,z}^*+ib_{s,z}^*)e^{iqa/4}
+ (a_{s,z}^*-ib_{s,z}^*)e^{iqa/4} \nonumber \\ && \
+ (a_{s,z}^*-ib_{s,z}^*)e^{iqa/4}
- (a_{s,z}^*+ib_{s,z}^*)e^{iqa/4} \nonumber \\ && \
- (a_{s,z}^*+ib_{s,z}^*)e^{-iqa/4}
+ (a_{s,z}^*-ib_{s,z}^*)e^{-iqa/4} ] \nonumber \\
&=& 4 a_{cz} b_{s,z}^* i \left[ - e^{-iqa/4} - e^{iqa/4} \right]
\nonumber \\ &=& -8i a_{cz} b_{s,z}^* \cos(aq/4) \ ,
\end{eqnarray}
so that
\begin{eqnarray}
V_y(L_{zz}^y) &=& 16 N_{uc} Q_y L_{zz}^y \cos(qa/4) \Im [ a_{cz} b_{s,z}^* ] \ .
\end{eqnarray}

We consider the terms proportional to $L_{xy}^y$:
\begin{eqnarray}
V_y(L_{xy}^y) &=& Q_y L_{xy}^y \biggl( [ S_x(5) - S_x(6)]
\nonumber \\ && \ \times [S_y(1) + S_y(2) - S_y(3) - S_y(4)]
\nonumber \\ && \ + [ S_y(5) - S_y(6)] \nonumber \\ 
&& \ \times [S_x(1) + S_x(2) - S_x(3) - S_x(4)]
\biggr) \nonumber \\ & \equiv&
N_{uc}Q_y L_{xy}^y [u_{xy} + u_{xy}^*] \ ,
\end{eqnarray}
where
\begin{eqnarray}
u_{xy} &=& b_{cx} \left[ (-ia_{s,y}^*+b_{s,y}^*)e^{-iqa/4}
+ (-ia_{s,y}^*-b_{s,y}^*)e^{-iqa/4} \right. \nonumber \\ && \
- (ia_{s,y}^*+b_{s,y}^*)e^{iqa/4}
- (ia_{s,y}^*-b_{s,y}^*)e^{iqa/4} \nonumber \\ && \
+ (-ia_{s,y}^*+b_{s,y}^*)e^{iqa/4}
+ (-ia_{s,y}^*-b_{s,y}^*)e^{iqa/4} \nonumber \\ && \
\left. - (ia_{s,y}^*+b_{s,y}^*)e^{-iqa/4}
- (ia_{s,y}^*-b_{s,y}^*)e^{-iqa/4} \right] \nonumber \\ &&
+ a_{cy} \left[
  (a_{s,x}^* - i b_{s,x}^*) e^{-iqa/4} \right. \nonumber \\ && \
+ (-a_{s,x}^* - i b_{s,x}^*) e^{-iqa/4}
- (a_{s,x}^* + i b_{s,x}^*) e^{iqa/4} \nonumber \\ && \
- (-a_{s,x}^* + i b_{s,x}^*) e^{iqa/4} \nonumber \\ && \
+ (a_{s,x}^* - i b_{s,x}^*) e^{iqa/4}
+ (-a_{s,x}^* - i b_{s,x}^*) e^{iqa/4} \nonumber \\ && \
\left. - (a_{s,x}^* + i b_{s,x}^*) e^{-iqa/4}
- (-a_{s,x}^* + i b_{s,x}^*) e^{-iqa/4} \right] \nonumber \\ &=&
4 i b_{cx} \left[ -a_{s,y}^* e^{-iqa/4} - a_{s,y}^* e^{iqa/4} \right]
\nonumber \\ && \
+ 4 i a_{cy} \left[ -b_{s,x}^*e^{-iqa/4} -b_{s,x}^*e^{iqa/4} \right]
\nonumber \\ &=&  -8i[b_{cx} a_{s,y}^* +a_{cy} b_{s,x}^*] \cos (qa/4) \ ,
\end{eqnarray}
so that
\begin{eqnarray}
V_y(L_{xy}^y) &=& 16 N_{uc} Q_y L_{xy}^y \cos(qa/4) \nonumber \\ && 
\ \times \Im \left[ b_{cx}a_{s,y}^* + a_{cy} b_{s,x}^* \right] \ .
\end{eqnarray}

We consider the terms proportional to $L_{xz}^y$:
\begin{eqnarray}
V_y(L_{xz}^y) &=& Q_y L_{xz}^y \Biggl( [ S_x(5) + S_x(6)] \nonumber \\
&& \ \times  [S_z(1) + S_z(2) - S_z(3) - S_z(4)]
\nonumber \\ && \
+ [ S_z(5) + S_z(6)] \nonumber \\ && \ \times
[S_x(1) + S_x(2) - S_x(3) - S_x(4)]
\Biggr) \nonumber \\ & \equiv&
N_{uc}Q_y L_{xz}^y [u_{xz} + u_{xz}^*] \ ,
\end{eqnarray}
where
\begin{eqnarray}
u_{xz} &=& b_{cx} \left[ (a_{s,z}^*-ib_{s,z}^*)e^{-iqa/4}
+ (a_{s,z}^*+ib_{s,z}^*)e^{-iqa/4} \right. \nonumber \\ && \
- (a_{s,z}^*+ib_{s,z}^*)e^{iqa/4}
- (a_{s,z}^*-ib_{s,z}^*)e^{iqa/4} \nonumber \\ && \
- (a_{s,z}^*-ib_{s,z}^*)e^{iqa/4}
- (a_{s,z}^*+ib_{s,z}^*)e^{iqa/4} \nonumber \\ && \
\left. + (a_{s,z}^*+ib_{s,z}^*)e^{-iqa/4}
+ (a_{s,z}^*-ib_{s,z}^*)e^{-iqa/4} \right] \nonumber \\ && \
+ a_{cz} \left[ (a_{s,x}^* -i b_{s,x}^*) e^{-iqa/4} 
\right. \nonumber \\ && \
+ (-a_{s,x}^* -i b_{s,x}^*) e^{-iqa/4}  \nonumber \\ && \
- (a_{s,x}^* +i b_{s,x}^*) e^{iqa/4} 
- (-a_{s,x}^* +i b_{s,x}^*) e^{iqa/4} \nonumber \\ && \ 
+ (a_{s,x}^* -i b_{s,x}^*) e^{iqa/4} 
+ (-a_{s,x}^* -i b_{s,x}^*) e^{iqa/4}  \nonumber \\ && \
\left. - (a_{s,x}^* +i b_{s,x}^*) e^{-iqa/4} 
- (-a_{s,x}^* +i b_{s,x}^*) e^{-iqa/4} \right] \nonumber \\ &=& 
4 b_{cx} a_{s,z}^* [e^{-iqa/4} - e^{-iqa/4} ] \nonumber \\ && \
+ 4 a_{cz} b_{s,x}^* [-ie^{-iqa/4} - i e^{iqa/4} ]
\nonumber \\ &=&
-8ib_{cx} a_{s,z}^* \sin (qa/4) -8i a_{cz} b_{s,x}^* \cos (qa/4) \ ,
\end{eqnarray}
so that
\begin{eqnarray}
V_y (L_{xz}^y) &=& 16 N_{uc} Q_y L_{xz}^y \Biggl(
\sin(qa/4) \Im[ b_{cx}a_{s,z}^*] \nonumber \\ && \
+ \cos(qa/4) \Im[ a_{cz} b_{s,x}^*] \Biggr) \ .
\end{eqnarray}

We consider the terms proportional to $L_{yz}^y$:
\begin{eqnarray}
V_y(L_{yz}^y) &=& Q_y L_{yz}^y \Biggl( [ S_y(5) - S_y(6)] \nonumber \\
&& \ \times [S_z(1) - S_z(2) - S_z(3) + S_z(4)]
\nonumber \\ && \ + [ S_z(5) - S_z(6)] \nonumber \\ && \
\times [S_y(1) - S_y(2) - S_y(3) + S_y(4)] \Biggr)
\nonumber \\ &\equiv& N_{uc}Q_y L_{yz}^y [u_{yz} + u_{yz}^*] \ ,
\end{eqnarray}
where
\begin{eqnarray}
u_{yz} &=& a_{cy} \left[ (a_{s,z}^*-ib_{s,z}^*)e^{-iqa/4}
- (a_{s,z}^*+ib_{s,z}^*)e^{-iqa/4} \right. \nonumber \\ && \
- (a_{s,z}^*+ib_{s,z}^*)e^{iqa/4}
+ (a_{s,z}^*-ib_{s,z}^*)e^{iqa/4} \nonumber \\ && \
+ (a_{s,z}^*-ib_{s,z}^*)e^{iqa/4}
- (a_{s,z}^*+ib_{s,z}^*)e^{iqa/4} \nonumber \\ && \
\left. - (a_{s,z}^*+ib_{s,z}^*)e^{-iqa/4}
+ (a_{s,z}^*-ib_{s,z}^*)e^{-iqa/4} \right] \nonumber \\ && \
+ a_{c,z} \left[ (-ia_{s,y}^*+b_{s,y}^*)e^{-iqa/4} \right.
\nonumber \\ && \ - (-ia_{s,y}^*-b_{s,y}^*)e^{-iqa/4} 
\nonumber \\ && \ - (ia_{s,y}^*+b_{s,y}^*)e^{iqa/4}
+ (ia_{s,y}^*-b_{s,y}^*)e^{iqa/4} \nonumber \\ && \
- (-ia_{s,y}^*+b_{s,z}^*)e^{iqa/4}
+ (-ia_{s,y}^*-b_{s,y}^*)e^{iqa/4} \nonumber \\ && \
\left. + (ia_{s,y}^*+b_{s,y}^*)e^{-iqa/4}
- (ia_{s,y}^*-b_{s,y}^*)e^{-iqa/4} \right] \nonumber \\ &=&
-4i a_{cy} b_{sz}^* \left[ e^{-iqa/4} + e^{iqa/4} \right] 
\nonumber \\ && \ +
4 a_{cz} b_{s,y}^* \left[ e^{-iqa/4} - e^{iqa/4} \right] \nonumber \\
&=& -8i a_{cy} b_{s,z}^* \cos (aq/4) -8ia_{cz} b_{s,y}^*
\sin (qa/4) \ ,
\end{eqnarray}
so that
\begin{eqnarray}
V_y(L_{yz}^y) &=& 16 N_{uc} Q_y L_{yz}^y \left( \cos(qa/4) \Im[ a_{cy}b_{s,z}^*] 
\right. \nonumber \\ && \left. \ + \sin(qa/4) \Im[a_{cz} b_{s,y}^* ] \right) \ .
\end{eqnarray}

We consider the terms proportional to $F_x^y$:
\begin{eqnarray}
V_y(F_x^y) &=& Q_y F_x^y \Biggl( [ S_y(5) - S_y(6)] \nonumber \\ &&
\ \times [S_z(1) - S_z(2) - S_z(3) + S_z(4)]
\nonumber \\ && \ - [ S_z(5) - S_z(6)] \nonumber \\ &&
\ \times [S_y(1) - S_y(2) - S_y(3) + S_y(4)] \Biggr)
\nonumber \\  &\equiv& N_{uc}Q_y F_x^y [u_x + u_x^*] \ ,
\end{eqnarray}
where
\begin{eqnarray}
u_x &=& a_{cy} \left[ (a_{s,z}^*-ib_{s,z}^*)e^{-iqa/4}
- (a_{s,z}^*+ib_{s,z}^*)e^{-iqa/4} \right. \nonumber \\ && \
- (a_{s,z}^*+ib_{s,z}^*)e^{iqa/4}
+ (a_{s,z}^*-ib_{s,z}^*)e^{iqa/4} \nonumber \\ && \
+ (a_{s,z}^*-ib_{s,z}^*)e^{iqa/4}
- (a_{s,z}^*+ib_{s,z}^*)e^{iqa/4} \nonumber \\ && \
\left. - (a_{s,z}^*+ib_{s,z}^*)e^{-iqa/4}
+ (a_{s,z}^*-ib_{s,z}^*)e^{-iqa/4} \right] \nonumber \\ && \
+ a_{c,z} \left[ -(-ia_{s,y}^*+b_{s,y}^*)e^{-iqa/4}
\right. \nonumber \\ && \
+ (-ia_{s,y}^*-b_{s,y}^*)e^{-iqa/4} \nonumber \\ && \
+ (ia_{s,y}^*+b_{s,y}^*)e^{iqa/4}
- (ia_{s,y}^*-b_{s,y}^*)e^{iqa/4} \nonumber \\ && \
+ (-ia_{s,y}^*+b_{s,y}^*)e^{iqa/4}
- (-ia_{s,y}^*-b_{s,y}^*)e^{iqa/4} \nonumber \\ && \
\left. - (ia_{s,y}^*+b_{s,y}^*)e^{-iqa/4}
+ (ia_{s,y}^*-b_{s,y}^*)e^{-iqa/4} \right] \nonumber \\ &=& \
-4i a_{cy} b_{s,z}^* \left[ e^{-iqa/4} + e^{iqa/4} \right] 
\nonumber \\ && \ + 
4 a_{cz} b_{s,y}^* \left[ e^{iqa/4} - e^{-iqa/4} \right] \nonumber \\
&=& -8ia_{cy} b_{s,z}^* \cos (qa/4) + 8i a_{cz} b_{s,y}^* \sin (qa/4) \ ,
\end{eqnarray}
so that
\begin{eqnarray}
V_y(F_x^y) &=& 16 N_{uc} Q_y F_x^y \Biggl(
\cos(qa/4) \Im[ a_{cy}b_{s,z}^*] \nonumber \\ && \
+ \sin (qa/4) \Im [a_{cz}^* b_{s,y} ] \Biggr)  \ .
\end{eqnarray}

We consider the terms proportional to $F_y^y$:
\begin{eqnarray}
V_y(F_y^y) &=& Q_y F_y^y \Biggl( -[ S_x(5) + S_x(6)] \nonumber \\ &&
\ \times [S_z(1) + S_z(2) - S_z(3) - S_z(4)]
\nonumber \\ && \
+ [ S_z(5) + S_z(6)] \nonumber \\ && \
\times [S_x(1) + S_x(2) - S_x(3) - S_x(4)] \Biggr) \nonumber
\\ &\equiv& N_{uc}Q_y L_{yz}^y [u_y + u_y^*] \ ,
\end{eqnarray}
where
\begin{eqnarray}
u_y &=& b_{cx} \left[ -(a_{s,z}^*-ib_{s,z}^*)e^{-iqa/4}
- (a_{s,z}^*+ib_{s,z}^*)e^{-iqa/4} \right. \nonumber \\ && \
+ (a_{s,z}^*+ib_{s,z}^*)e^{iqa/4}
+ (a_{s,z}^*-ib_{s,z}^*)e^{iqa/4} \nonumber \\ && \
+ (a_{s,z}^*-ib_{s,z}^*)e^{iqa/4}
+ (a_{s,z}^*+ib_{s,z}^*)e^{iqa/4} \nonumber \\ && \
\left. - (a_{s,z}^*+ib_{s,z}^*)e^{-iqa/4}
- (a_{s,z}^*-ib_{s,z}^*)e^{-iqa/4} \right] \nonumber \\ && \
+ a_{c,z} \left[ (a_{s,x}^* -ib_{s,x}^*)e^{-iqa/4}
\right. \nonumber \\ && \
+ (-a_{s,x}^*-ib_{s,x}^*)e^{-iqa/4} \nonumber \\ && \
- (a_{s,x}^*+ib_{s,x}^*)e^{iqa/4}
- (-a_{s,x}^*+ib_{s,x}^*)e^{iqa/4} \nonumber \\ && \
(a_{s,x}^*-ib_{s,x}^*)e^{iqa/4}
+ (-a_{s,x}^*-ib_{s,x}^*)e^{iqa/4} \nonumber \\ && \
\left. - (a_{s,x}^*+ib_{s,x}^*)e^{-iqa/4}
- (-a_{s,x}^*+ib_{s,x}^*)e^{-iqa/4} \right] \nonumber \\ &=& \
4b_{cx} a_{s,z}^* \left[ -e^{-iqa/4} + e^{iqa/4} \right] \nonumber
\\ && \ + 4 a_{cz} b_{s,x}^* \left[ -i e^{-iqa/4} -i e^{iqa/4} \right] \nonumber \\
&=& 8i b_{cx} a_{s,z}^* \sin (qa/4) -8i a_{cz} b_{s,x}^* \cos (aq/4) \ ,
\end{eqnarray}
so that
\begin{eqnarray}
V_y(F_y^y) &=& 16 N_{uc} Q_y F_y^y \Biggl( \sin(qa/4) \Im[ b_{cx}^* a_{s,z}]
\nonumber \\ && \ + \cos(aq/4) \Im [ a_{cz} b_{s,x}^*] \Biggr)  \ .
\end{eqnarray}

Finally, we consider the terms proportional to $F_z^y$:
\begin{eqnarray}
V_y(F_z^y) &=& Q_y F_z^y \Biggl( [ S_x(5) - S_x(6)] \nonumber \\ && \
\times [S_y(1) + S_y(2) - S_y(3) - S_y(4)] \nonumber \\ && \
- [ S_y(5) - S_y(6)] \nonumber \\ && \
\times [S_x(1) + S_x(2) - S_x(3) - S_x(4)] \Biggr)
\nonumber \\ &\equiv& N_{uc}Q_y F_z^y [u_z + u_z^*] \ ,
\end{eqnarray}
where
\begin{eqnarray}
u_z &=& b_{cx} \left[ (-ia_{s,y}^*+b_{s,y}^*)e^{-iqa/4}
\right. \nonumber \\ && \
+ (-ia_{s,y}^*-b_{s,y}^*)e^{-iqa/4} \nonumber \\ && \
- (ia_{s,y}^*+b_{s,y}^*)e^{iqa/4}
- (ia_{s,y}^*-b_{s,y}^*)e^{iqa/4} \nonumber \\ && \
+ (-ia_{s,y}^*+b_{s,y}^*)e^{iqa/4}
+ (-ia_{s,y}^*-b_{s,y}^*)e^{iqa/4} \nonumber \\ && \
\left. - (ia_{s,y}^*+b_{s,y}^*)e^{-iqa/4}
- (ia_{s,y}^*-b_{s,y}^*)e^{-iqa/4} \right] \nonumber \\ && \
+ a_{c,y} \left[ -(a_{s,x}^*-ib_{s,x}^*)e^{-iqa/4}
\right. \nonumber \\ && \
- (-a_{s,x}^*-ib_{s,x}^*)e^{-iqa/4} \nonumber \\ && \
+ (a_{s,x}^*+ib_{s,x}^*)e^{iqa/4}
+ (-a_{s,x}^*+ib_{s,x}^*)e^{iqa/4} \nonumber \\ && \
- (a_{s,x}^*-ib_{s,x}^*)e^{iqa/4}
- (-a_{s,x}^*-ib_{s,x}^*)e^{iqa/4} \nonumber \\ && \
\left. + (a_{s,x}^*+ib_{s,x}^*)e^{-iqa/4}
+ (-a_{s,x}^*+ib_{s,x}^*)e^{-iqa/4}\right] \nonumber \\ &=& \
4ib_{cx} a_{s,y}^* \left[ -e^{-iqa/4} - e^{iqa/4} \right] \nonumber \\ && \
+4 i a_{cy} b_{s,x}^* \left[ e^{-iqa/4} + e^{iqa/4} \right] \nonumber \\
&=& -8i b_{cx} a_{s,y}^* \cos(qa/4) + 8 i  a_{cy} b_{s,x}^* \cos(qa/4) \ ,
\end{eqnarray}
so that
\begin{eqnarray}
V_y(F_z^y)  &=& 16 N_{uc} Q_y F_z^y \cos(qa/4) \nonumber \\ &&
\ \times \Im[ b_{cx}a_{s,y}^* + a_{cy}^* b_{s,x} ] \ .
\end{eqnarray}
\end{appendix}

\end{document}